\documentclass[english,superscriptaddress,floatfix,preprintnumbers,nofootinbib]{revtex4}
\usepackage[T1]{fontenc}
\usepackage[utf8]{inputenc}
\usepackage{xcolor}
\definecolor{BLUE}{rgb}{0, 0, 1}
\definecolor{RED}{rgb}{1, 0, 0}
\definecolor{blu}{rgb}{0, 0, 0.699219}
\definecolor{bluc}{rgb}{0, 0, 0.898438}
\definecolor{blus}{rgb}{0, 0, 0.398438}
\definecolor{lime}{rgb}{0.652344, 0.808594, 0.226562}
\definecolor{rosso}{rgb}{0.601562, 0, 0}
\definecolor{rossoc}{rgb}{0.800781, 0, 0}
\definecolor{rossos}{rgb}{0.449219, 0, 0}
\definecolor{verde}{rgb}{0.0585938, 0.75, 0.308594}
\definecolor{verdec}{rgb}{0.0664062, 0.851562, 0.347656}
\definecolor{verdes}{rgb}{0.046875, 0.601562, 0.246094}
\usepackage{babel}
\usepackage{mathtools}
\usepackage{amsmath}
\usepackage{amsthm}
\usepackage{amssymb}
\usepackage{graphicx}
\usepackage{wasysym}
\usepackage[bookmarks=false, breaklinks=false,pdfborder={0 0 1},backref=section,colorlinks=false]{hyperref}
\hypersetup{citecolor=purple,linkcolor=blue,linkcolor=blu,pdfstartview=FitH,urlcolor=rossos,citecolor=rossos}
\usepackage{babel}
\usepackage{amsthm}
\usepackage{wasysym}

\usepackage{xcolor}
\usepackage[flushleft]{threeparttable}
\usepackage[justification=raggedright, margin=5mm, font=sf, small, labelfont=bf, format=plain]{caption}
\usepackage{amsthm}
\usepackage{wasysym}
\usepackage{color}
\usepackage{textcomp}
\usepackage{amsfonts}
\usepackage{slashed}
\usepackage{adjustbox}
\usepackage{lipsum}
\usepackage{rotating}
\usepackage{xcolor}
\usepackage{wrapfig}
\usepackage{epsfig}
\usepackage{tikzsymbols}
\usepackage{tikz}
\usepackage{tabularx}
\usepackage{ulem}
\usepackage{enumerate}
\usepackage{txfonts}
\DeclareRobustCommand{\orcidicon}{
	\begin{tikzpicture}
	\draw[lime, fill=lime] (0, 0)
	circle [radius=0.2]
	node[white] {{\fontfamily{qag}\selectfont \tiny ID}};
	\draw[white, fill=white] (-0.0625, 0.095)
	circle [radius=0.007];
	\end{tikzpicture}
	\hspace{-2mm}}
\foreach \x in {A, ..., Z}{\expandafter\xdef\csname orcid\x\endcsname{\noexpand\href{https://orcid.org/\csname orcidauthor\x\endcsname}
			{\noexpand\orcidicon}}}

\begin{document}
\title{\textcolor{blu}{Phenomenology of the Minimal Scale Invariant Two-Higgs-Doublet Model}}

\author{Nabil Baouche\orcidA{}}
\email{nabil.baouche@univ-jijel.dz}
\affiliation{Faculty of Science and Technology, University of Jijel PB 98 Ouled
Aissa, DZ-18000 Jijel, Algeria.}
\affiliation{Laboratoire de Physique des Particules et Physique Statistique, Ecole
Normale Superieure, BP 92 Vieux Kouba, DZ-16050 Algiers, Algeria.}

\author{Amine Ahriche\orcidB{}}
\email{ahriche@sharjah.ac.ae}
\affiliation{Department of Applied Physics and Astronomy, University of Sharjah,
P.O. Box 27272 Sharjah, UAE.}
\affiliation{Research Institute of Sciences and Engineering, University of Sharjah,
P.O. Box 27272 Sharjah, UAE.}
\affiliation{Laboratoire de Physique des Particules et Physique Statistique, Ecole
Normale Superieure, BP 92 Vieux Kouba, DZ-16050 Algiers, Algeria.}

\begin{abstract}
We perform a comprehensive phenomenological analysis of the Scale Invariant Two-Higgs-Doublet Model (\textit{SI2HDM}). In this minimal framework, electroweak symmetry breaking is triggered radiatively without the need for an additional real singlet scalar. After imposing stringent theoretical and experimental constraints, including electroweak precision tests and $B^0_d \to \mu^+\mu^-$ flavor bounds, we identify a highly constrained viable parameter space that accommodates two distinct scenarios: the light-dilaton ($50~\textrm{GeV} \le m_\eta < m_h$) and purely radiative Higgs mass (PRHM, $m_\eta > m_h$) scenarios. Within this viable space, the charged Higgs mass is constrained to $140 \le m_{H^\pm} \le 600~\textrm{GeV}$ while the CP-odd scalar mass spans $400 \le m_{A^0} \le 600~\textrm{GeV}$. This tight parameter space renders the model highly predictive: radiative corrections suppress the trilinear Higgs self-couplings, reducing the di-Higgs production cross section at the 13~TeV LHC by $55\%-70\%$ in the light-dilaton scenario and up to $60\%$ in the PRHM scenario compared to the Standard Model prediction. These results highlight the predictive power and testability of the \textit{SI2HDM} as a minimal framework for radiatively driven electroweak symmetry breaking.
\end{abstract}
\maketitle

\section{Introduction~\label{sec:Intro}}

The discovery of the Higgs boson in July 2012 by the ATLAS and CMS
collaborations at the LHC~\cite{ATLAS:2012yve,CMS:2012qbp} provided
experimental confirmation of the Higgs mechanism for electroweak symmetry
breaking. While this achievement represents a major milestone in elementary
particle physics, several fundamental questions remain unresolved,
such as the baryon asymmetry of the universe, the origin and smallness
of neutrino masses, the nature of dark matter (DM), and the hierarchy
problem. These issues motivate us to go beyond the Standard Model
(BSM), which can be extended in many ways. In particular, the hierarchy
problem, associated with the sensitivity of the Higgs mass to high-energy
physics, provides a fundamental theoretical motivation for investigating
BSM scenarios.

A popular approach to addressing the hierarchy problem is to impose
classical scale invariance (SI) by setting the Higgs mass parameter
to zero ($\mu^{2}=0$). In this framework, electroweak symmetry breaking
(EWSB) is not triggered by a negative mass term but instead occurs
radiatively through a mechanism known as dimensional transmutation,
where quantum corrections break classical scale invariance~\cite{Coleman:1973jx}.
The spontaneous breaking of scale invariance is associated with a
massless tree-level pseudo-Goldstone boson, or ``dilaton'', which
acquires a mass radiatively after EWSB. The realization of EWSB through
the Coleman-Weinberg (CW) mechanism in the standard model (SM) and
its extensions has been extensively studied in the literature (e.g.,
see~\cite{Alexander-Nunneley:2010tyr}).

Many SI extensions of the SM address the hierarchy problem, as well
as other issues such as DM and neutrino oscillation data~\cite{Foot:2007ay,Karam:2015jta,Ahriche:2014oda,Lindner:2014oea,Humbert:2015epa,Ahriche:2016ixu,Ahriche:2016cio,Ahriche:2023hho,Nhi:2025iob,Chang:2007ki,Foot:2007as,Holthausen:2009uc,Shaposhnikov:2008xi,Goldberger:2007zk,Shaposhnikov:2009nk,Fuyuto:2015xsa,Meissner:2006zh,Foot:2007iy,Ghorbani:2022vtv,Kannike:2016wuy,Fuyuto:2015jha,Guo:2014bha,Ghorbani:2015xvz,Benic:2014aga,Ishiwata:2011aa,YaserAyazi:2018lrv,Cosme:2018nly,Karam:2016rsz,Khoze:2016zfi,YaserAyazi:2019caf,Marzo:2018nov,Marzola:2017jzl,Sannino:2015wka,Man:2013gpa,Apreda:2005yz,Arunasalam:2017ajm,Farzinnia:2014yqa,Lane:2018ycs,Ghorbani:2017lyk,Chiang:2017zbz}.
In many SI extensions, an additional real scalar singlet is introduced
to assist EWSB. Through its vacuum expectation value (VEV), this singlet
mixes with the SM Higgs doublet, resulting in two CP-even mass eigenstates,
one of which is massless at tree level (the dilaton). The light CP-even
scalar can be identified with the observed 125~\textrm{GeV} SM-like Higgs,
while the other CP-even scalar is heavier. This possibility defines
a purely radiative Higgs mass (PRHM) scenario, whose conditions have
been detailed in~\cite{Ahriche:2021frb}. In this scenario, radiative
corrections lift the tree-level mass of the dilaton to the observed
125~\textrm{GeV} Higgs mass, while the scalar doublet-singlet mixing is brought
into agreement with experimental data. The PRHM scenario has been
investigated in several contexts, including neutrino mass generation,
DM~\cite{Soualah:2021xbn}, and gravitational waves (GW) from phase
transitions~\cite{Ahriche:2023jdq}.

There are different approaches to achieving radiative EWSB in classically
scale-invariant models, where the tree-level mass parameters vanish
and the scale is generated by dimensional transmutation à la Coleman-Weinberg~\cite{Coleman:1973jx}.
The best-known approach is the Gildener-Weinberg (GW) formalism~\cite{Gildener:1976ih},
which assumes that the tree-level potential has a flat direction $\hat{n}$
in field space, along which the quartic couplings satisfy $\lambda_{ijkl}\,\hat{n}_i \hat{n}_{j}\hat{n}_{k}\hat{n}_{l}=0$
at some renormalization scale $\Lambda_{{\rm GW}}$. Along this direction,
the one-loop effective potential takes the form 
\begin{equation}
V(\varphi)=A\,\varphi^{4}+B\,\varphi^{4}\ln\frac{\varphi^{2}}{\Lambda^{2}_{{\rm GW}}},
\end{equation}
whose minimum fixes the VEV, while the loop corrections are treated
as a small perturbation. The scalar along the flat direction is then
a pseudo-Goldstone boson of scale invariance, the dilaton, with a
loop-suppressed mass 
\begin{equation}
m^{2}_{D}=\frac{1}{8\pi^{2}v^{2}_{\varphi}}\sum_i n_i \,m^{4}_i .
\end{equation}
In this approach, the 125~\textrm{GeV} Higgs corresponds to the heavier, tree-level
massive eigenstate, whose mixing with the dilaton is fixed at leading
order by the flat direction. A more general approach was proposed
in Ref.~\cite{Ahriche:2021frb}, which does not rely on expanding
around a tree-level flat direction, but instead uses the full one-loop
effective potential $V(h,\phi)$ with renormalization conditions (in
a modified $\overline{{\rm DR}}$ scheme) that fix the tadpoles and
the Higgs mass at the vacuum. Consequently, the loop contributions
to the scalar mass matrix are not required to be small. This allows
for a second regime in which the light CP-even state, massless at
tree level, becomes the observed 125~\textrm{GeV} Higgs through radiative
corrections, accompanied by a heavier singlet-like scalar $S$. This
defines the PRHM scenario, while the conventional GW light-dilaton
scenario is recovered when the loop corrections are small compared
with the tree-level mass of the heavier state. In the PRHM regime,
radiative corrections also bring the scalar mixing into agreement
with the LHC signal-strength constraints and generate the trilinear
couplings $\lambda_{HHH}$ and $\lambda_{HHS}$, which vanish at tree
level. It is worth noting that this greater generality concerns the
regime covered rather than the loop order: both approaches rely on
the same one-loop effective potential. Therefore, in the regime of
sizable loop corrections, the results should be understood as one-loop
estimates, subject to the usual renormalization-scale and scheme dependence.

In SI models where EWSB is assisted by a real singlet, the singlet
VEV provides an additional degree of freedom and plays an important
role in satisfying experimental constraints. This raises the question
of whether an additional real singlet scalar is necessary to achieve
radiative EWSB and realize the PRHM scenario. In this work, we address
this question by considering an SI model in which EWSB is achieved
without an additional real singlet scalar ~\cite{Lee:2012jn}. This
model, which we call the SI two-Higgs-doublet model (\textit{SI2HDM}),
is the usual 2HDM ~\cite{Branco:2011iw,Lee:1973iz,Gunion:2002zf}
with all quadratic terms in the scalar potential set to zero, where
EWSB is achieved radiatively and both Higgs doublets acquire VEVs.

Here, we investigate the viability of this model and examine how the
absence of an additional singlet affects its viable parameter space.
In addition, we explore the possibility of realizing the PRHM scenario
within this framework. After applying theoretical and experimental
constraints, we delineate the viable parameter space and identify
two distinct scenarios: a PRHM scenario $m_{h}<m_{\eta}$ and a dilaton
scenario $m_{\eta}<m_{h}$. We find that the charged Higgs mass is
constrained to lie below approximately 600~\textrm{GeV}, while the CP-odd
scalar mass lies within the range $400<m_{A^{0}}<600$~\textrm{GeV}. Notably,
radiative corrections suppress the trilinear Higgs coupling and reduce
the di-Higgs production cross section at the 13~TeV LHC by up to
70\% in the dilaton scenario and 60\% in the PRHM scenario. These
results highlight the predictive power and testability of the \textit{SI2HDM},
offering insights into the hierarchy problem while remaining compatible
with precision measurements.

This work is organized as follows. Section~\ref{sec:Mod} outlines
the \textit{SI2HDM} model and its mass spectrum. Section~\ref{sec:Theo}
examines the theoretical and experimental constraints on the model.
In Section~\ref{sec:collider}, we investigate the distinctive collider
phenomenology of the \textit{SI2HDM}, with a focus on di-Higgs production.
Section~\ref{sec:Resu} presents the numerical results, maps the
viable parameter space, and emphasizes the distinctive phenomenological
features of the model. Finally, Section~\ref{sec:Conc} concludes
with a discussion of the broader implications of the \textit{SI2HDM}
and possible avenues for experimental verification at future colliders.

\section{Model and Mass Spectrum~\label{sec:Mod}}

In this study, we adopt the model presented in Ref.~\cite{Lee:2012jn}.
This model extends the SM particle content by introducing a second
(SU(2)L) Higgs doublet. Classical scale invariance is imposed by excluding
all dimensionful quadratic terms from the scalar potential. The most
general renormalizable and gauge-invariant scalar potential at tree
level in this framework is given by 
\begin{equation}
\begin{array}{c}
V^{0}=\frac{1}{6}\lambda_1 \left(\Phi^{\dagger}_1 \Phi_1 \right)^{2}+\frac{1}{6}\lambda_2 \left(\Phi^{\dagger}_2 \Phi_2 \right)^{2}+\lambda_{3}\left(\Phi^{\dagger}_1 \Phi_1 \right)\left(\Phi^{\dagger}_2 \Phi_2 \right)+\lambda_{4}|\Phi^{\dagger}_1 \Phi_2 |^{2}\\
+\left\{\frac{1}{2}\lambda_{5}\left(\Phi^{\dagger}_1 \Phi_2 \right)^{2}+\lambda_{6}\left(\Phi^{\dagger}_1 \Phi_1 \right)\left(\Phi^{\dagger}_1 \Phi_2 \right)+\lambda_{7}\left(\Phi^{\dagger}_2 \Phi_2 \right)\left(\Phi^{\dagger}_2 \Phi_1 \right)+h.c.\right\} ,
\end{array}\label{poten}
\end{equation}
where $\lambda_{i=1,7}$ are dimensionless scalar couplings. To suppress
tree-level Higgs mediated flavor changing neutral currents (FCNC)
in the Yukawa sector~\cite{Glashow:1976nt}, we impose a discrete
$Z_2 $ symmetry under which the Higgs doublets transform as $\left(\Phi_1 \to-\Phi_1 ,\Phi_2 \to+\Phi_2 \right)$.
This symmetry forces the quartic couplings $\lambda_{6}$ and $\lambda_{7}$
to vanish. In this scenario, $\lambda_{5}$ can be chosen to be real
without loss of generality, which precludes explicit $CP$ violation
at tree level. Furthermore, we assume that $CP$ is not spontaneously
broken. Consequently, only the CP-even neutral components of the two
Higgs doublets $\Phi_{1,2}$ acquire VEVs, denoted $\upsilon_{1,2}$.
The doublets can therefore be parameterized as $\Phi^{T}_i =\Big(\begin{array}{cc}
\phi^{+}_i , & (\upsilon_i +\phi_i +ia_i )/\sqrt{2}\end{array}\Big)$, where $\upsilon=\sqrt{\upsilon^{2}_1 +\upsilon^{2}_2}=246~\textrm{~\textrm{GeV}}$
is the SM electroweak VEV; and the ratio of the VEVs is defined as
$\tan\beta=t_{\beta}=\upsilon_2 /\upsilon_1 $.

In our setup, we consider an \textit{SI2HDM} with a type-I Yukawa
structure. Under the imposed $Z_2 $ symmetry, the most general Yukawa
Lagrangian is given by: 
\begin{equation}
\mathcal{L}_{Yukawa}\supset-y_{t}\Big(\begin{array}{cc}
\bar{t}_{L} & \bar{b}_{L}\end{array}\Big)\Big(i\sigma_2 \Phi^{*}_2 \Big)t_{R}-y_{b}\Big(\begin{array}{cc}
\bar{t}_{L} & \bar{b}_{L}\end{array}\Big)\Phi_2 b_{R}-y_{\tau}\Big(\begin{array}{cc}
\bar{\nu}_{L} & \bar{\tau}_{L}\end{array}\Big)\Phi_2 \tau_{R}+h.c.,
\end{equation}
where all SM fermions couple to the second doublet $\Phi_2 $. Consequently,
fermion masses receive contributions exclusively from $\Phi_2 $.
This leads to the relation $y_{f}=\frac{\sqrt{2}m_{f}}{s_{\beta}\upsilon}$,
where $c_{\beta}=\cos\beta$ and $s_{\beta}=\sin\beta$.

The tree-level tadpole conditions, $\partial V_{0}/\partial\phi_1 =\partial V_{0}/\partial\phi_2 =0$,
yield the relations $\lambda_1 =-3\lambda_{345}t^{2}_{\beta}$ and
$\lambda_2 =-3\lambda_{345}/t^{2}_{\beta}$, with $\lambda_{345}=\lambda_{3}+\lambda_{4}+\lambda_{5}$.

The model contains eight scalar degrees of freedom (dofs): the two
CP-even dofs $\phi_{1,2}$ mix to form two CP-even mass eigenstates
$(h,\eta)$, where $h$ is the SM-like Higgs boson and $\eta$ is
a second scalar that can be either lighter or heavier than the SM-like
Higgs boson. The two CP-odd dofs $a^{0}_{1,2}$ give rise to the neutral
Goldstone boson $\chi^{0}$ and a CP-odd mass eigenstate $A^{0}$.
Finally, the four charged dofs $\phi^{\pm}_{1,2}$ yield the charged
Goldstone boson $\chi^{\pm}$ and a charged mass eigenstate $H^{\pm}$.

At tree level, the mass matrices in the bases ($\big(\phi_1 ,\phi_2 \big)$,
$\big(a^{0}_1 ,a^{0}_2 \big)$, and $\big(\phi^{\pm}_1 ,\phi^{\pm}_2 \big)$)
share a common form: 
\begin{equation}
\mathcal{M}^{2}_{Q}=m^{2}_{Q}\left(\begin{array}{cc}
s^{2}_{\beta}, & -s_{\beta}c_{\beta}\\
-s_{\beta}c_{\beta}, & c^{2}_{\beta}
\end{array}\right),\label{eq:M0}
\end{equation}
which yields eigenvalues of $0$ and $m^{2}_{Q}$. The massless eigenstate
corresponds to the dilaton (in the CP-even sector), the neutral Goldstone
boson (in the CP-odd sector), and the charged Goldstone boson (in
the charged sector). The massive eigenstate corresponds to the Higgs
boson $h$, the CP-odd scalar $A^{0}$, and the charged scalar $H^{\pm}$,
respectively. Consequently, the quartic couplings can be expressed
in terms of tree-level masses as $\lambda_{3}=\big(2m^{2}_{H^{\pm}}-m^{2}_{h}\big)/\upsilon^{2}$,
$\lambda_{4}=\big(m^{2}_{A^{0}}-2m^{2}_{H^{\pm}}\big)/\upsilon^{2}$,
and $\lambda_{5}=-m^{2}_{A^{0}}/\upsilon^{2}$.

The one-loop effective scalar potential can be written as a function
of the CP-even, CP-odd, and charged scalar fields as: 
\begin{equation}
\begin{array}{c}
V^{1-\ell}(\Phi_{1,2})=V^{0}(\Phi_{1,2})+V^{CT}(\Phi_{1,2})+\sum_i n_i G\Big(m^{2}_i (\Phi_{1,2})\Big),\end{array}\label{eq:V1l}
\end{equation}
where $V^{CT}(\Phi_{1,2})=V^{0}(\Phi_{1,2},\lambda_i \to\delta\lambda_i )$
is the counterterm potential, $n_i $ are the dofs multiplicities\footnote{We have $n_{h}=n_{A^{0}}=1$, $n_{H^{\pm}}=2$, $n_{W}=6$, $n_{Z}=3$
and $n_{t}=n_{b}=-12$.}, and $m^{2}_i (\Phi_{1,2})$ are the field dependent masses. The
function $G(x)=\frac{x^{2}}{64\pi^{2}}\Big(\log\big(\frac{x}{\Lambda^{2}}\big)-\frac{3}{2}\Big)$
is computed within the $\overline{DR}$ scheme~\cite{Martin:2001vx},
and $\Lambda$ is the renormalization scale. The introduction of the
dimensionful parameter $\Lambda$ explicitly breaks the classical
scale invariance of the model. For phenomenological applications,
such as collider physics analyses, this scale is conventionally fixed
at the electroweak scale, identified with the Higgs mass $\Lambda=m_{h}=125.18~\textrm{~\textrm{GeV}}$.

By applying the one-loop tadpole conditions, $\partial V^{1-\ell}(\phi_1 ,\phi_2 )/\partial\phi_1 =\partial V^{1-\ell}(\phi_1 ,\phi_2 )/\partial\phi_2 =0$,
at the vacuum expectation values ($\langle\phi_i \rangle=\upsilon_i $,
$\langle a_i \rangle=\langle\phi^{\pm}_i \rangle=0$), the counter
terms $\delta\lambda_1 $ and $\delta\lambda_2 $ can be expressed
in terms of $\delta\lambda_{345}$ and specific combinations of the
field dependent masses $m^{2}_i (\Phi_{1,2})$ and their derivatives,
as detailed in Appendix~\ref{sec:Mass}. Consequently, the one-loop
scalar mass matrices can be written as: 
\begin{equation}
\begin{array}{cc}
\Big(\mathcal{M}^{2}_{Q}\Big)^{1-\ell}= & \left(m^{2}_{Q}-\delta\lambda_{Q}\upsilon^{2}\right)\left(\begin{array}{cc}
s^{2}_{\beta}, & -s_{\beta}c_{\beta}\\
-s_{\beta}c_{\beta}, & c^{2}_{\beta}
\end{array}\right)+m^{2}_{Q}\left(\begin{array}{cc}
a_{Q} & c_{Q}\\
c_{Q} & b_{Q}
\end{array}\right),\end{array}\label{M1-1}
\end{equation}
where the sets $\{\mathcal{M}^{2}_{Q},\delta\lambda_{Q},m^{2}_{Q},a_{Q},b_{Q},c_{Q}\}=$$\{\mathcal{M}^{2}_{0^{+}},\delta\lambda_{345},m^{2}_{h},a_{0^{+}},b_{0^{+}},c_{0^{+}}\}$,
$\{\mathcal{M}^{2}_{0^{-}},\delta\lambda_{5},m^{2}_{A^{0}},a_{0^{-}},b_{0^{-}},c_{0^{-}}\}$
and $\{\mathcal{M}^{2}_{\pm},\frac{1}{2}\delta\lambda_{45},m^{2}_{H^{\pm}},a_{\pm},b_{\pm},c_{\pm}\}$
for the CP-even, CP-odd and charged sectors, respectively. The parameters
$a_{Q},~b_{Q}$, and $c_{Q}$ are defined in Appendix~\ref{sec:Mass}.
In Refs.~\cite{Lee:2012jn,Lane:2018ycs,Lane:2019dbc,Eichten:2022vys},
it has been argued that a well chosen value for the quartic coupling
counterterms can make the $2\times2$ CP-odd (charged) squared mass
matrix having same value at both tree- and one-loop levels, which
means that the mass and mixing values at one-loop should have the
same tree-level values, in addition, the Goldstones are automatically
massless. In Appendix~\ref{sec:Gold}, we will show that this is
wrong and the choice of the counter term values of the quartic coupling
should be made based on physical requirements.

To ensure a physically consistent spectrum, the counterterms are chosen
such that one CP-even mass eigenstate reproduces the measured Higgs
mass, while the Goldstone bosons in the CP-odd and charged sectors
remain massless. This requirement yields the following expressions
for the counterterms: 
\begin{align*}
\delta\lambda_{Q} & =\frac{m^{2}_{Q}}{\upsilon^{2}}\frac{a_{Q}s^{2}_{\beta}+b_{Q}c^{2}_{\beta}+c^{2}_{Q}-a_{Q}b_{Q}-2c_{Q}s_{\beta}c_{\beta}}{1-a_{Q}c^{2}_{\beta}-b_{Q}s^{2}_{\beta}-2c_{Q}s_{\beta}c_{\beta}},
\end{align*}
and the effective mixing angle at the one-loop level for all sectors
is given by: 
\begin{align}
s_{Q} & =-sign\big[\big(\mathcal{M}^{2}_{Q}\big)^{1-\ell}_{12}\big]\sqrt{\frac{1}{2}+\frac{1}{2}\frac{\big(\mathcal{M}^{2}_{Q}\big)^{1-\ell}_{11}-\big(\mathcal{M}^{2}_{Q}\big)^{1-\ell}_{22}}{\sqrt{\big[\big(\mathcal{M}^{2}_{Q}\big)^{1-\ell}_{11}-\big(\mathcal{M}^{2}_{Q}\big)^{1-\ell}_{22}\big]^{2}+4\big[\big(\mathcal{M}^{2}_{Q}\big)^{1-\ell}_{12}\big]^{2}}}}.\label{eq:sb}
\end{align}

However, for the CP-even sector in the PRHM case, the relevant definition
is instead 
\begin{align}
s_{ev} & =sign\big[\big(\mathcal{M}^{2}_{0^{+}}\big)^{1-\ell}_{12}\big]\sqrt{\frac{1}{2}-\frac{1}{2}\frac{\big(\mathcal{M}^{2}_{0^{+}}\big)^{1-\ell}_{11}-\big(\mathcal{M}^{2}_{0^{+}}\big)^{1-\ell}_{22}}{\sqrt{\big[\big(\mathcal{M}^{2}_{0^{+}}\big)^{1-\ell}_{11}-\big(\mathcal{M}^{2}_{0^{+}}\big)^{1-\ell}_{22}\big]^{2}+4\big[\big(\mathcal{M}^{2}_{0^{+}}\big)^{1-\ell}_{12}\big]^{2}}}}.
\end{align}

The one-loop masses for CP-odd and charged scalars are $\Big(m^{2}_{A^{0}}\Big)^{1-\ell}=m^{2}_{A^{0}}(1+a_{0^{-}}+b_{0^{-}})-\delta\lambda_{5}\upsilon^{2}$
and $\Big(m^{2}_{H^{\pm}}\Big)^{1-\ell}=m^{2}_{H^{\pm}}(1+a_{\pm}+b_{\pm})-\frac{1}{2}\delta\lambda_{45}\upsilon^{2}$.
To ensure a light dilaton scenario, i.e., the SM-like Higgs to be
the heavier CP-even eigenstate, where $m_{\eta}<m_{h}$, the condition
$m^{2}_{h}<\Big(\mathcal{M}^{2}_{0^{+}}\Big)^{1-\ell}_{22}+\Big(\mathcal{M}^{2}_{0^{+}}\Big)^{1-\ell}_{11}<2~m^{2}_{h}$
must be satisfied. This condition translates to $\frac{\upsilon^{2}}{m^{2}_{h}}\delta\lambda_{345}<a_{0^{+}}+b_{0^{+}}<1+\frac{\upsilon^{2}}{m^{2}_{h}}\delta\lambda_{345}$.
Conversely, in the PRHM case, where the Higgs corresponds to the light
CP-even state ($m_{\eta}>m_{h}$), the condition becomes $\Big(\mathcal{M}^{2}_{0^{+}}\Big)^{1-\ell}_{22}+\Big(\mathcal{M}^{2}_{0^{+}}\Big)^{1-\ell}_{11}>2~m^{2}_{h}$,
which implies $a_{0^{+}}+b_{0^{+}}>1+\frac{\upsilon^{2}}{m^{2}_{h}}\delta\lambda_{345}$~\cite{Ahriche:2021frb}.

Since the one-loop mixing angles defined in Eq.~(\ref{eq:sb}) generally
differ from their tree-level value ($s_{\beta}$), as do all model
parameters, it is useful to define the deviation $\Delta_{\mathcal{O}}=\mathcal{O}^{1-\ell}-\mathcal{O}^{tree}$
for the observables $\mathcal{O}=\{s_{ev}=\sin\beta_{even},s_{od}=\sin\beta_{odd},s_{ch}=\sin\beta_{charged}\}$,
while the relative difference $\delta_{\mathcal{O}}=\frac{\mathcal{O}^{1-\ell}-\mathcal{O}^{tree}}{\mathcal{O}^{tree}}$
provides a useful measure for all masses and couplings.

We note that the renormalization approach used here, as in~\cite{Ahriche:2021frb},
is the most general one, since the counterterms are defined using physical
measurements. In addition, the mass of the extra CP-even scalar (whether
a dilaton or a heavy scalar) and the scalar mixing are determined exactly.
The dilaton mass value\footnote{In the GW approach.~\cite{Lane:2019dbc,Eichten:2022vys} where the
light CP-even mass that is radiatively generated corresponds to the
125~\textrm{GeV} SM-like Higgs, implies a sum rule $(\sum_i n_i m^{4}_i )^{1/4}>540~\textrm{~\textrm{GeV}}$~\cite{Lane:2019dbc,Eichten:2022vys}.
Here, $i$ runs over all BSM fields, which in our case are $H^{\pm}$,
$A^{0}$, and $\eta$. This is consistent with our results, as the extra CP-even scalar
is heavier than the 125~\textrm{GeV} SM-like Higgs, as shown in the numerical
results section. According
to our numerical results $(\sum_i n_i m^{4}_i )^{1/4}\sim~208-675~\textrm{~\textrm{GeV}}$,
which means that the choice $(\sum_i n_i m^{4}_i )^{1/4}>540~\textrm{~\textrm{GeV}}$~\cite{Lane:2019dbc,Eichten:2022vys}
discards a significant part of the viable parameter space.} is generally estimated using~\cite{Hempfling:1996ht},
where the positivity of the squared mass here implies the vacuum stability
discussed below in Eq.~(\ref{eq:vs1loop}).

\section{Theoretical \& Experimental Constraints~\label{sec:Theo}}

In our study, we impose the following theoretical and experimental
constraints: perturbativity, vacuum stability, perturbative unitarity,
electroweak precision tests (EWPT), di-photon Higgs decay, LEP direct
searches for charginos and neutralinos, and constraints on the Higgs
signal strength modifiers. In what follows, we discuss each constraint
in detail.

\vspace{0.6cm}
 \textbf{Perturbativity:}\\
 Perturbativity is a fundamental requirement that ensures the validity
of perturbative calculations. For the quartic couplings in the SM
and its extensions, this requires the couplings to remain sufficiently
small to avoid strong coupling or non-perturbative effects. Therefore,
all physical quartic couplings must satisfy: $\lambda_{1,}\lambda_{2,}\left|\lambda_{3}+\lambda_{4}\right|,\left|\lambda_{3}+\lambda_{4}\pm\lambda_{5}\right|<4\pi.$

\vspace{0.6cm}
 \textbf{Vacuum stability:}\\

To ensure the scalar potential is bounded from below, we derive necessary
stability conditions. These apply in all field directions where quartic
terms dominate. First, in both the $h-\eta$ and $\chi^{0}-A^{0}$
field planes, the potential must satisfy~\cite{Klimenko:1984qx}
$\lambda_{1,}\lambda_{2,}\lambda_{3}+\sqrt{\lambda_1 \lambda_2}>0,\lambda_{345}+\sqrt{\lambda_1 \lambda_2}>0$.
Second, in the $\chi^{\pm}-H^{\pm}$ plane, we obtain the additional
constraint $\lambda_{3}+\lambda_{4}+\frac{2}{3}\sqrt{\lambda_1 \lambda_2}>0$.
The leading term in the effective potential in Eq.~(\ref{eq:V1l})
is proportional to $\varphi^{4}\log\varphi$, rather than $\varphi^{4}$,
where $\varphi$ denotes an arbitrary direction in the $\{\phi_1 ,\phi_2 \}$
field space. Consequently, the one-loop vacuum stability conditions
follow from the positivity of the coefficients of the $\varphi^{4}\log\varphi$
terms in Eq.~(\ref{eq:V1l}). This means that if the field dependent
masses can be written in the CP-even field space as $m^{2}_i =\frac{1}{2}(\alpha_i \phi^{2}_1 +\beta_i \phi^{2}_2 )$,
then these conditions are written as~\cite{Ahriche:2021frb} 
\begin{equation}
\sum_i n_i \alpha^{2}_i ,\sum_i n_i \beta^{2}_i >0.\label{eq:vs1loop}
\end{equation}

\vspace{0.6cm}
 \textbf{Perturbative unitarity:}\\
 To derive unitarity constraints at high energies, one must compute
the decay amplitudes for scalar-scalar scattering and ensure they
respect tree-level unitarity~\cite{Cornwall:1974km}. The dominant
contributions to these amplitudes are mediated exclusively by the
quartic couplings~\cite{Arhrib:2000is}. The full scattering amplitude
matrix can be decomposed into submatrices based on the symmetry of
the initial/final state of the processes $S_1 S_2 \to S_{3}S_{4}$.
In the \textit{SI2HDM}, the global $Z_2 $ symmetry and electric
charge are exact symmetries. Consequently, we identify two big matrices
according to the electric charge of the initial/final states: a $14\times14$
matrix in the basis $\{\frac{\phi_1 \phi_1}{\sqrt{2}},\frac{\phi_2 \phi_2}{\sqrt{2}},\frac{a_1 a_1}{\sqrt{2}},\frac{a_2 a_2}{\sqrt{2}},\phi^{+}_1 \phi^{-}_1 ,\phi^{+}_2 \phi^{-}_2 ,\phi_1 \phi_2 ,a_1 a_2 ,\phi^{+}_1 \phi^{-}_2 ,\phi^{+}_2 \phi^{-}_1 ,\phi_1 a_2 ,\phi_2 a_1 ,\phi_1 a_1 ,\phi_2 a_2 \}$,
representing the neutral initial/final states. Due to the global $Z_2 $
symmetry, it can be divided into 3 independent submatrices as shown
in Appendix~\ref{sec:Matrice}. The charged initial/final states
can be described by the $8\times8$ matrix that is defined in the
basis $\left\{\phi_1 \phi^{+}_1 ,\phi_2 \phi^{+}_2 ,a_1 \phi^{+}_1 ,a_2 \phi^{+}_2 ,\phi_1 \phi^{+}_2 ,\phi_2 \phi^{+}_1 ,a_1 \phi^{+}_2 ,a_2 \phi^{+}_1 \right\} $.
Because of the global $Z_2 $ symmetry, this matrix can be simplified
into two independent $4\times4$ matrices. The theory respects unitarity
if each eigenvalue of the scattering matrices shown in Appendix~\ref{sec:Matrice}
does not exceed $8\pi$.

\vspace{0.6cm}
 \textbf{Electroweak precision tests:}\\
 The oblique parameters $S$, $T$, and $U$ provide a robust framework
for assessing the compatibility of BSM theories with high precision
electroweak measurements. These parameters characterize the impact
of new physics (NP), mediated via loop corrections, on the electroweak
gauge boson propagators. In extensions of the SM that introduce additional
scalar fields mixing with the Higgs boson after EWSB, gauge interactions
induce modifications to the $W$ and $Z$ boson self energies. Such
corrections shift the oblique parameters, providing measurable deviations
from SM predictions. The scalar interactions given in Eq.~(\ref{eq:V1l})
contribute to the oblique parameters~\cite{Grimus:2008nb}. The oblique
parameters in the \textit{SI2HDM} are given by 
\begin{align}
\Delta S & =\frac{1}{24\pi}\left\{\left[2s^{2}_{W}-1\right]^{2}G\left(m^{2}_{H^{\pm}},m^{2}_{H^{\pm}},m^{2}_{Z}\right)+c^{2}_{\beta_{od}-\beta_{ev}}\left[G\left(m^{2}_{h},m^{2}_{A^{0}},m^{2}_{Z}\right)+\hat{G}\left(m^{2}_{\eta},m^{2}_{Z}\right)\right]\right.\nonumber \\
 & \left.+s^{2}_{\beta_{ev}-\beta_{od}}\left[G\left(m^{2}_{\eta},m^{2}_{A^{0}},m^{2}_{Z}\right)+\hat{G}\left(m^{2}_{h},m^{2}_{Z}\right)\right]-\hat{G}(m^{2}_{h},m^{2}_{Z})+\log\frac{m^{2}_{\eta}}{m^{2}_{H^{\pm}}}+\log\frac{m^{2}_{A^{0}}}{m^{2}_{H^{\pm}}}\right\} ,\\
\Delta T & =\frac{1}{16~\pi~s^{2}_{W}~m^{2}_{W}}\left\{c^{2}_{\beta_{ch}-\beta_{ev}}F\left(m^{2}_{H^{\pm}},m^{2}_{h}\right)+s^{2}_{\beta_{ev}-\beta_{ch}}F\left(m^{2}_{H^{\pm}},m^{2}_{\eta}\right)+c^{2}_{\beta_{ch}-\beta_{od}}F\left(m^{2}_{H^{\pm}},m^{2}_{A^{0}}\right)\right.\nonumber \\
 & -s^{2}_{\beta_{ev}-\beta_{od}}~F\left(m^{2}_{\eta},m^{2}_{A^{0}}\right)-3\left\{F\left(m^{2}_{Z},m^{2}_{h}\right)-F\left(m^{2}_{W},m^{2}_{h}\right)\right\} -c^{2}_{\beta_{od}-\beta_{ev}}~F\left(m^{2}_{h},m^{2}_{A^{0}}\right)\nonumber \\
 & \left.+3c^{2}_{\beta_{od}-\beta_{ev}}~\left\{F\left(m^{2}_{Z},m^{2}_{\eta}\right)-F\left(m^{2}_{W},m^{2}_{\eta}\right)\right\} +3s^{2}_{\beta_{ev}-\beta_{od}}\left\{F\left(m^{2}_{Z},m^{2}_{h}\right)-F\left(m^{2}_{W},m^{2}_{h}\right)\right\} \right\} ,
\end{align}
\begin{align}
\Delta U & =\frac{1}{24\pi}\left\{c^{2}_{\beta_{ch}-\beta_{ev}}G\left(m^{2}_{H^{\pm}},m^{2}_{h},m^{2}_{w}\right)+s^{2}_{\beta_{ev}-\beta_{ch}}G\left(m^{2}_{H^{\pm}},m^{2}_{\eta},m^{2}_{w}\right)+c^{2}_{\beta_{ch}-\beta_{od}}G\left(m^{2}_{H^{\pm}},m^{2}_{A^{0}},m^{2}_{w}\right)\right.\nonumber \\
 & +s^{2}_{\beta_{ev}-\beta_{od}}\left[\hat{G}(m^{2}_{h},m^{2}_{w})-\hat{G}(m^{2}_{h},m^{2}_{Z})\right]+c^{2}_{\beta_{od}-\beta_{ev}}\left[\hat{G}(m^{2}_{\eta},m^{2}_{w})-\hat{G}(m^{2}_{\eta},m^{2}_{Z})\right]+\hat{G}(m^{2}_{h},m^{2}_{Z})-\hat{G}(m^{2}_{h},m^{2}_{w})\nonumber \\
 & \left.-c^{2}_{\beta_{od}-\beta_{ev}}G\left(m^{2}_{h},m^{2}_{A^{0}},m^{2}_{Z}\right)-s^{2}_{\beta_{ev}-\beta_{od}}G\left(m^{2}_{\eta},m^{2}_{A^{0}},m^{2}_{Z}\right)-\left[2s^{2}_{w}-1\right]^{2}G\left(m^{2}_{H^{\pm}},m^{2}_{H^{\pm}},m^{2}_{Z}\right)\right\} ,
\end{align}
where $c_{\beta_i -\beta_{j}}=\cos(\beta_i -\beta_{j})$ and $s_{\beta_i -\beta_{j}}=\sin(\beta_i -\beta_{j})$,
with $i,j=ev,od,ch.$ Here, $s_{W}=\sin\theta_{W}$ and $c_{W}=\cos\theta_{W}$,
with $\theta_{W}$ being the Weinberg mixing angle. The functions
$F(x,y)$, $G(x,y,z)$, and $\hat{G}(x,y,z)$ are loop integrals provided
in the literature~\cite{Grimus:2008nb}.

To ensure consistency between theoretical predictions and experimental
measurements of the electroweak oblique parameters, we define a chi-squared
function, $\chi^{2}_{EWPT}$, as follows: 
\begin{equation}
\chi^{2}_{EWPT}=\sum_{\mathcal{O}=S,T,U}\frac{(\Delta\mathcal{O}-\Delta\mathcal{O}^{exp})^{2}}{\sigma^{2}_{\mathcal{O}}}<3.53,
\end{equation}
where the experimental central values and associated uncertainties
for the oblique parameters are $\Delta S=-0.04\pm0.1,\;\Delta T=0.01\pm0.12,\:\Delta U=-0.01\pm0.09$~\cite{ParticleDataGroup:2024cfk}.

\vspace{0.6cm}
 \textbf{Undetermined Higgs decay:}\\
 The Higgs boson can decay into a pair of new scalars $S=\eta,A^{0}$
if $m_{S}<m_{h}/2$. The decay width for this process is given by
\begin{equation}
\varGamma~(h\to SS)=\Theta(m_{h}-2m_{S})\frac{\lambda^{2}_{hSS}}{32\pi m_{h}}\sqrt{1-4\frac{m^{2}_{S}}{m^{2}_{h}}},\label{Gund}
\end{equation}
where $\lambda_{hSS}$ represents the scalar trilinear couplings provided
in Appendix~\ref{sec:Mass}. The undetermined Higgs decay width $\Gamma_{und}=\varGamma~(h\to\eta\eta)+\varGamma(h\to A^{0}A^{0})$
is distinct from the invisible decay width, as the light scalar(s)
($S=\eta,A^{0}$) may decay to light fermions ($S\to f\bar{f}$) and
could be detected by future experiments. These decays do not correspond
to known SM processes; hence, the process $h\to\eta\eta\to f_1 \bar{f_1}f_2 \bar{f_2}$
is termed unconstrained. Therefore, the total Higgs decay width can
be expressed as 
\begin{equation}
\Gamma^{tot}_{h}=\varGamma_{und}+\Gamma^{SM}_{h}\underset{X=SM}{\sum}\kappa^{2}_{X}~B(h\to XX)_{SM},\label{decayT}
\end{equation}
where $\ensuremath{\Gamma}^{SM}_{h}=4.02~\mathrm{MeV}$~\cite{Heinemeyer:2013tqa}
is the total Higgs decay width in the SM, and $\kappa_{V}=s_{\beta-\beta_{ev}}$,
~$\left(\kappa_{F}=\frac{c_{\beta_{ev}}}{s_{\beta}}\right)$ represent
the Higgs coupling modifiers to pairs of gauge bosons $(V)$ (fermions
$(f)$), respectively, relative to their SM predictions. Here, the
dilaton ($\eta$) has couplings similar to the SM Higgs but modified
by the factors $\zeta_{V}=c_{\beta-\beta_{ev}}$ and $\zeta_{F}=\frac{s_{\beta_{ev}}}{s_{\beta}}$.
It is also necessary to properly define the CP-odd $(A^{0})$ scalar
coupling modifiers to pairs of gauge bosons $(V)$ and fermions $(f)$,
which are represented by the symbols $\varrho_{V}=0$ and $\varrho_{F}=\frac{c_{\beta_{od}}}{s_{\beta}}$.
The undetermined branching ratio must satisfy the constraint $\mathcal{B}_{und}<0.22$~\cite{CMS:2015chx}.
Additionally, the total Higgs decay width must lie within the range:
$\Gamma_{h}=3.7^{+1.9}_{-1.4}~\mathrm{MeV}$~\cite{ParticleDataGroup:2024cfk}.

\vspace{0.6cm}
 \textbf{The di-photon Higgs decay:}\\
 In the \textit{SI2HDM}, the presence of the
charged scalar $H^{\pm}$ modifies the Higgs decay branching ratio
to diphotons, $\mathcal{B}\left(h\to\gamma\gamma\right)$. This modification
is constrained experimentally by the measured signal strength $\mu_{\gamma\gamma}=1.10\pm0.06$,
derived from a combined ATLAS and CMS analysis~\cite{ParticleDataGroup:2024cfk}.
Within the \textit{SI2HDM} framework, we evaluate the signal strength
for $h\to\gamma\gamma$ as 
\begin{align}
\mu_{XX} & =\frac{[\sigma(pp\to h)\times\mathcal{B}(h\to\gamma\gamma)]^{SI2HDM}}{[\sigma(pp\to h)\times\mathcal{B}(h\to\gamma\gamma)]^{SM}}\nonumber \\
 & =\kappa^{2}_{F}\frac{\Gamma^{SM}_{h}}{\Gamma^{tot}_{h}}\left|\frac{\frac{\lambda_{hH^{+}H^{-}}\upsilon}{2m^{2}_{H^{\pm}}}A^{\gamma\gamma}_{0}(\tau_{H^{\pm}})+\kappa_{V}A^{\gamma\gamma}_1 (\tau_{W})+\kappa_{F}\sum_{f}N^{f}Q^{2}_{e}A^{\gamma\gamma}_{1/2}(\tau_{f})}{A^{\gamma\gamma}_1 (\tau_{W})+\sum_{f}N^{f}Q^{2}_{e}A^{\gamma\gamma}_{1/2}(\tau_{f})}\right|^{2},
\end{align}
with $\sigma(pp\to h)^{SI2HDM}$($\sigma(pp\to h)^{SM}$) is the gluon-gluon
fusion (ggF) production cross section in the \textit{SI2HDM} (SM)
model and $\mathcal{B}_{BSM}$ represents the Higgs undetermined branching
ratio, $\lambda_{hH^{+}H^{-}}$ is the one-loop scalar trilinear coupling
defined in Appendix~\ref{sec:Mass}, and $N^{f}$ and $Q^{f}_{e}$
are the color factor and electric charge of the fermions, respectively.
The functions $A^{\gamma\gamma}_{0,1/2,1}$ are given in~\cite{Djouadi:2005gi}.

\vspace{0.6cm}
 \textbf{Gauge boson decay widths:}\\
 In many SM extensions, such as the \textit{SI2HDM} or other scenarios
with additional gauge bosons or scalars, the decays of $W$ and $Z$
gauge bosons into new invisible or exotic particles must be carefully
constrained to avoid conflicts with experimental observations. Therefore,
invisible decays of the $W$ and $Z$ gauge bosons via processes such
as $W^{\pm}\to A^{0}W^{\pm}$ and $Z\to H^{+}H^{-}$ are excluded
to preserve the unmodified $W$ and $Z$ gauge boson decay modes.
Thus, we impose the conditions: $\min(m_{\eta}+m_{A^{0}},2m_{H^{\pm}})>m_{Z}$,
and $\min(m_{\eta}+m_{H^{\pm}},m_{H^{\pm}}+m_{A^{0}})>m_{W}$.

\vspace{0.6cm}
 \textbf{LEP direct searches for charginos and neutralinos:}\\
 Searches for charginos and neutralinos at LEP-II~\cite{LEPHiggsWorkingGroupforHiggsbosonsearches:2001ogs}
set experimental lower limits on scalar masses, requiring $m_{H^{\pm}}>78~\textrm{~\textrm{GeV}}$.

\vspace{0.6cm}
 \textbf{Higgs signal strength modifiers:}\\
 The discovery of the Higgs boson by the ATLAS and CMS collaborations
at the LHC has opened a new window for probing BSM physics~\cite{ATLAS:2012yve,CMS:2012qbp}.
These experimental results have enabled high precision measurements
of multiple Higgs boson coupling strengths. This precision allows
stringent constraints on SM extensions, particularly models involving
new scalar fields that couple to the SM Higgs doublet through mixing
mechanisms, as implemented in the \textit{SI2HDM}. The signal strength
modifier is an experimentally measured quantity for the combined production
and decay process, defined as the ratio of the measured Higgs boson
rate to its SM prediction. For a specific Higgs boson production channel
and decay into a particular final state, the signal strength is expressed
as 
\begin{equation}
\mu_{XX}=\frac{\sigma(pp\to h\to XX)^{SI2HDM}}{\sigma(pp\to h\to XX)_{SM}}=\frac{\sigma(pp\to h)^{SI2HDM}\times B(h\to XX)^{SI2HDM}}{\sigma(pp\to h)_{SM}\times B(h\to XX)_{SM}},
\end{equation}
where $\sigma(pp\to h)$ denotes the Higgs production cross section,
and $B(h\to XX)$ is the branching ratio into the final state $XX$.
The partial Higgs signal strength modifier for the channel $h\to XX$
can be simplified as 
\begin{equation}
\mu_{XX}=\kappa^{2}_{F}~\kappa^{2}_{X}\frac{\Gamma^{SM}_{h}}{\Gamma^{tot}_{h}},
\end{equation}
where $\kappa_{X}$ represents the normalized coupling strengths of
the Higgs boson to pairs of the final state $XX$, and $\Gamma^{tot}_{h}$
is the total Higgs decay width given in Eq.~(\ref{decayT}). To analyze
whether the model can provide a possible explanation for the observed
signal strengths in $\gamma\gamma$, $b\bar{b}$, $WW$, $ZZ$, $\tau^{+}\tau^{-}$,
and $\mu^{+}\mu^{-}$ channels, we perform a chi-squared analysis:
\begin{equation}
\chi^{2}_{SM}=\sum_{X}\frac{\left(\mu_{XX}-\mu^{exp}_{XX}\right)^{2}}{\left(\Delta\mu^{exp}_{XX}\right)^{2}},\label{eq:XSM}
\end{equation}
where $X=\gamma,b,Z,W,\tau,\mu$. In our numerical scan, we consider
the $\mu_{XX}$ results reported in~\cite{ParticleDataGroup:2024cfk}
and retain only BPs that correspond to a $1\sigma$ precision, i.e.,
$\chi^{2}_{SM}<7.04$.

\vspace{0.6cm}
 \textbf{Experimental flavor constraints:}\\

In our model, similarly to the Type-I 2HDM, only the second Higgs
doublet $\Phi_2 $ couples to all fermions, which leads to the interaction
vertex $H^{+}\bar{t}b$ being proportional to $\cot\beta$. This vertex
is flavor-changing charged current, analogous to the SM vertex $W^{+}\bar{t}b$,
meaning it does not introduce unwanted FCNC decays. Rather, it facilitates
allowed and expected flavor changing processes.

Confronting theoretical predictions with experimental results excludes
certain regions in the ($t_{\beta},m_{H^{\pm}}$) plane (for a summary,
see~\cite{Enomoto:2015wbn}). The relevant processes include hadronic
tau decays $\tau\to K\nu$ and $\tau\to\pi\nu$; the radiative $B$
meson decay $B\to X_{s}\gamma$; and the leptonic decays of mesons
such as $B\to\tau\nu$, $D,K\to\mu\nu$, $D_{s}\to\tau\nu,\mu\nu$,
$\pi\to\mu\nu$, and $B^{0}_{s,d}\to\mu^{+}\mu^{-}$. In our setup,
which resembles the Type-I 2HDM, the strongest constraint comes from
$B^{0}_{d}\to\mu^{+}\mu^{-}$~\cite{Enomoto:2015wbn}, which we include
in our numerical scan.

\section{Collider Phenomenology~\label{sec:collider}}

\begin{figure}[ht]
\centering \includegraphics[width=0.75\textwidth]{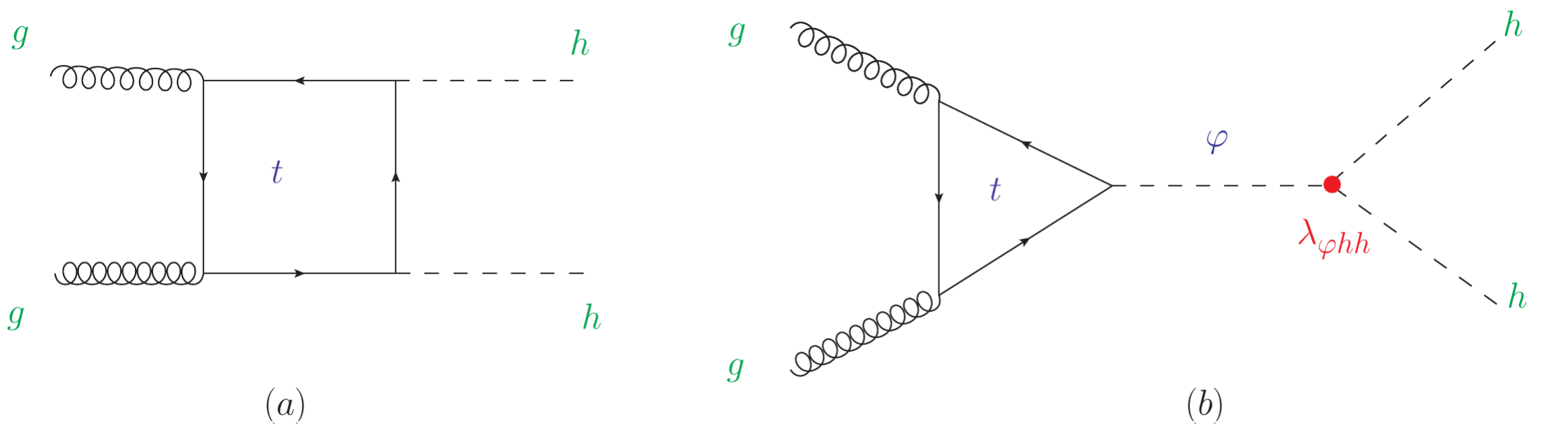} \caption{Feynman diagrams contributing to di-Higgs production via gluon fusion.
The left (right) diagram is referred to as the box (triangle) diagram
in the literature, where $\varphi\equiv h,\eta$.}
\label{diag} 
\end{figure}

At the LHC, the new neutral scalars ($\eta$ and $A^{0}$) can be
probed via the process $pp\to\eta/A^{0}\to f\bar{f},VV$. However,
their production cross section is suppressed due to the smallness
of the scalar couplings to quarks. In contrast, di-Higgs production
does not suffer from this coupling suppression, making it a pivotal
process in collider physics. It provides direct access to the Higgs
trilinear self coupling and serves as a stringent test of EWSB. While
the SM predicts a specific relationship between the Higgs self-coupling
and its mass, observed deviations could signal BSM physics. At lepton
colliders such as the proposed International Linear Collider (ILC),
the trilinear coupling can be accessed through double Higgs-strahlung
of $W$ or $Z$ bosons~\cite{Gounaris:1979px,Barger:1988jk,Ilyin:1995iy,Djouadi:1999gv,Tian:2010np},
$WW$/$ZZ$ fusion~\cite{Ilyin:1995iy,Djouadi:1999gv,Tian:2010np,Boudjema:1995cb,Barger:1988kb,Dobrovolskaya:1990kx,Abbasabadi:1988ja},
and gluon-gluon fusion (ggF) in $pp$ collisions at the LHC~\cite{Glover:1987nx,Plehn:1996wb,Dawson:1998py}.
The SM ggF cross section, suppressed by destructive interference between
box and triangle diagrams, is predicted to be $\sigma^{SM}_{ggF}=31.05^{+2.1}_{-7.2}\,\mathrm{fb}$~\cite{deFlorian:2019app}
at $\sqrt{s}=13\,\mathrm{TeV}$. However, in vector boson fusion (VBF)
mode at next-to-next-to-next-to-leading order, the cross section is
$\sigma^{SM}_{VBF}=1.73\pm0.04\,\mathrm{fb}$. Extensions of the Higgs
sector, such as the \textit{SI2HDM}, introduce new scalar states (the
CP-even scalar $\eta$) and modified coupling parameters that can
substantially alter the production dynamics. The \textit{SI2HDM} features
three primary contributions: two distinct triangle diagrams mediated
by the CP-even scalars $\varphi=h,\eta$, along with a box diagram
contribution, as shown in Fig.~\ref{diag}. Within the \textit{SI2HDM}
framework, additional contributions to di-Higgs production arise from
non-resonant Higgs production primarily via ggF at the LHC, followed
by Higgs decay. Moreover, modifications to the fermion Higgs couplings,
compared to the SM, could lead to significant deviations in the di-Higgs
production cross section. These effects may substantially enhance
the di-Higgs production rate, rendering it observable at the LHC or
at $e^{-}e^{+}$ colliders.

The di-Higgs production cross section in the SM has three contributions:
\begin{equation}
\sigma^{SM}(hh)=\sigma_{\Square}+\sigma_{\triangle}+\sigma_{\triangle\Square},
\end{equation}
which correspond to the box ($\sigma_{\Square}=65.0988\,\mathrm{fb}$),
triangle ($\sigma_{\triangle}=8.79372\,\mathrm{fb}$), and interference
terms ($\sigma_{\triangle\Square}=-42.84072\,\mathrm{fb}$), as obtained
using MadGraph~\cite{Alwall:2014hca}. In the \textit{SI2HDM}, the
di-Higgs production cross section can be expressed as~\cite{Ahriche:2014cpa,Baouche:2021wwa}
\begin{equation}
\sigma(hh)=\xi_1 \sigma_{\Square}+\xi_2 \sigma_{\triangle}+\xi_{3}\sigma_{\triangle\Square},
\end{equation}
where the SM corresponds to $\xi_1 =\xi_2 =\xi_{3}=1$. In the \textit{SI2HDM},
the coefficients $\xi_i $ are modified with respect to the SM as:
\begin{align}
\xi_1 =\kappa^{4}_{F},\,\xi_2  & =\left(\kappa_{F}\frac{\lambda_{hhh}}{\lambda^{SM}_{hhh}}+\zeta_{F}\frac{\lambda_{hh\eta}}{\lambda^{SM}_{hhh}}\frac{\hat{s}-m^{2}_{h}}{\hat{s}-m^{2}_{\eta}}\right)^{2},\,\xi_{3}=\kappa^{2}_{F}\,\left(\kappa_{F}\frac{\lambda_{hhh}}{\lambda^{SM}_{hhh}}+\zeta_{F}\frac{\lambda_{hh\eta}}{\lambda^{SM}_{hhh}}\frac{\hat{s}-m^{2}_{h}}{\hat{s}-m^{2}_{\eta}}\right),
\end{align}
where $\lambda_{hhh}$ and $\lambda_{hh\eta}$ are defined in Appendix~\ref{sec:Mass},
and $\kappa_{F}$ ($\zeta_{F}$) are the fermionic couplings for $h$
($\eta$) normalized to their SM values for Type I. The one-loop trilinear
Higgs coupling in the SM, $\lambda^{SM}_{hhh}$, is given by~\cite{Kanemura:2002vm}:
\begin{equation}
\lambda^{SM}_{hhh}\sim~eq\frac{3m^{2}_{h}}{\upsilon}\left[1-\frac{m^{4}_{t}}{\pi^{2}\upsilon^{2}m^{2}_{h}}\right].
\end{equation}

Here, $\sqrt{\hat{s}}$ is the partonic center-of-mass energy, which
corresponds to the invariant mass of the Higgs pair, $M_{hh}$. It
is much smaller than the hadronic collider energy $\sqrt{s}=13\,\mathrm{TeV}$.
Since $\sqrt{\hat{s}}$ is not fixed, we calculate the di-Higgs production
cross section using MadGraph5\_aMC@NLO~\cite{Alwall:2014hca} with
UFO files generated by FeynRules~\cite{Christensen:2008py}, which
properly handles the integration over $\sqrt{\hat{s}}$. The ATLAS
collaboration's combined statistical analysis of the full Run-II dataset
establishes an observed upper limit on the di-Higgs production cross
section of $2.9$ times the SM prediction, with an expected limit
of $2.4$ times at $95\%$ CL~\cite{ATLAS:2024ish}. These measurements
are highly valuable for constraining the scalar sector, particularly
in scenarios where the cross section exceeds SM predictions.

\section{Numerical Results~\label{sec:Resu}}

\begin{figure}[t]
\includegraphics[width=0.5\textwidth]{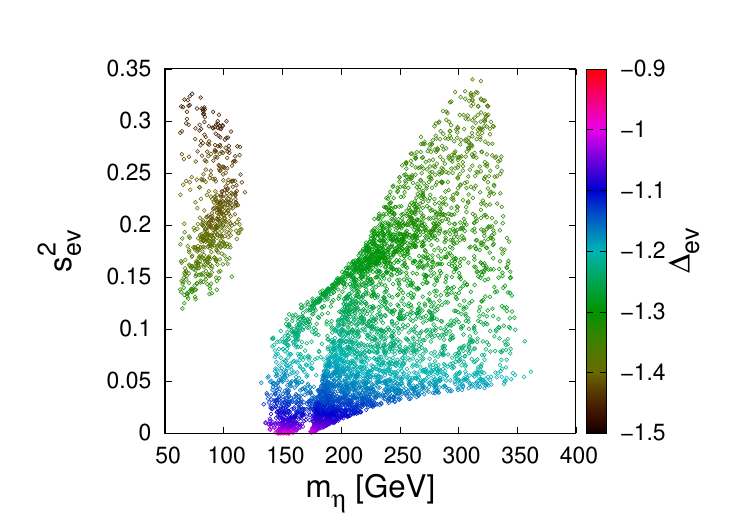}~\includegraphics[width=0.5\textwidth]{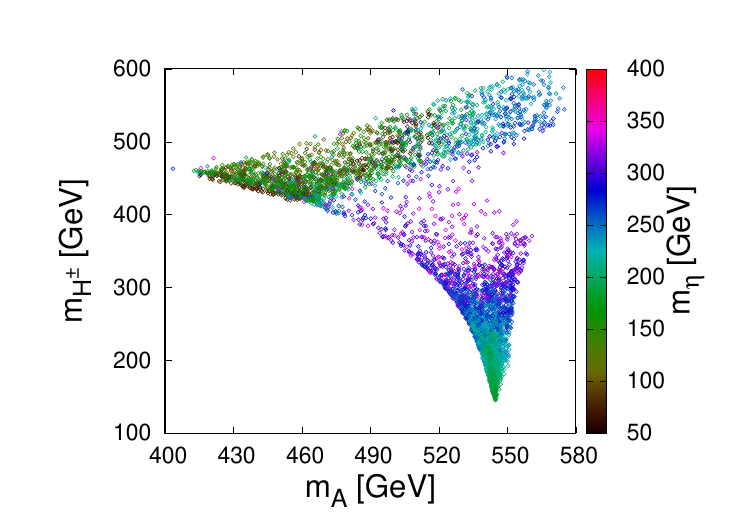}\\
 \includegraphics[width=0.5\textwidth]{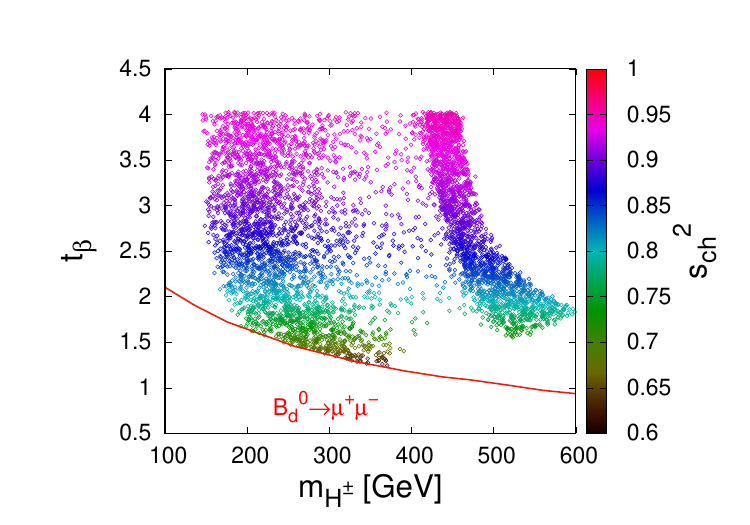}~\includegraphics[width=0.5\textwidth]{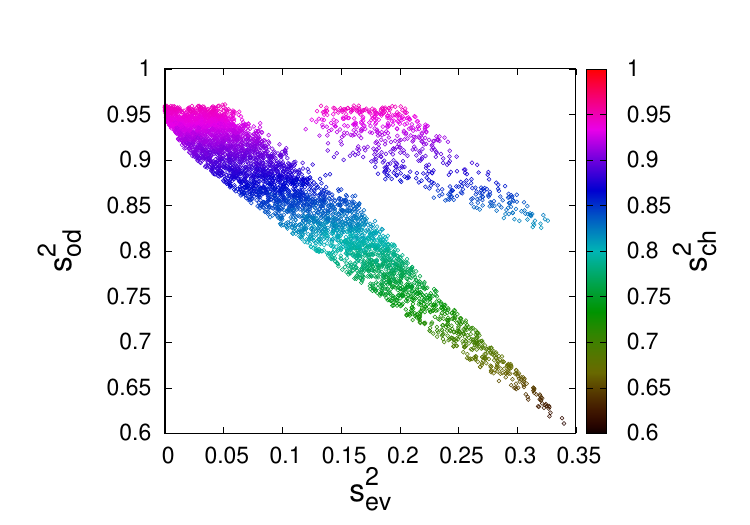}\\
 \caption{\textbf{Upper left panel}: The new CP-even scalar mass $m_{\eta}$
(in~\textrm{GeV}) versus the squared CP-even mixing $s^{2}_{ev}$, with the
color palette indicating the difference in CP-even mixing $s_{od}-s_{\beta}$.
\textbf{Upper right panel}: The CP-odd mass $m_{A^{0}}$ versus the
charged Higgs mass $m_{H^{\pm}}$ (both in~\textrm{GeV}), with the color palette
showing the new CP-even scalar mass $m_{\eta}$. \textbf{Lower left
panel}: The charged Higgs mass $m_{H^{\pm}}$ (in~\textrm{GeV}) versus the
tree-level mixing parameter $t_{\beta}$, with the color palette indicating
the squared charged mixing $s^{2}_{ch}$. The red dashed line represents
the experimental constraints from flavor physics that come from the
decay $B^{0}_{d}\to\mu^{+}\mu^{-}$~\cite{Enomoto:2015wbn}. \textbf{Lower
right panel}: The squared mixing $s^{2}_{od}$, $s^{2}_{ev}$ and
$s^{2}_{ch}$ for each sector.}
\label{Mass} 
\end{figure}

In this model, we have only three free parameters, so it is convenient
to choose them to be the mass of the CP-odd scalar $m_{A^{0}}$, the
charged scalar mass $m_{H^{\pm}}$, and the tree-level mixing angle
$s_{\beta}$, all at tree-level. In our analysis, we perform a numerical
scan over the parameter space within $m_{H^{\pm}}>78\,~\textrm{~\textrm{GeV}}$.
These parameter ranges are subject to all theoretical and experimental
constraints mentioned in the previous section. Our study examines
5000 BPs that comply with these constraints, which are shown in Fig.~\ref{Mass}.

The results shown in the upper left panel of Fig.~\ref{Mass} clearly
reveal two distinct regimes. The first corresponds to the light dilaton
scenario, characterized by $m_{\eta}<125\,~\textrm{GeV}$ and a CP-even mixing
in the range $s^{2}_{ev}\sim~eq0.10-0.35$, where the heavier CP-even
eigenstate is identified to be the SM-like Higgs boson. The second
is the PRHM scenario with $125\,~\textrm{GeV}<m_{\eta}$, in which the light
CP-even eigenstate is identified as the SM-like Higgs boson across
the entire scanned range of CP-even mixing. Its mixing structure is
governed by radiative effects, highlighting the dominant role of loop
corrections in determining the scalar mass spectrum. The upper right
panel of Fig.~\ref{Mass} shows the correlation between the mass
of the CP-odd scalar $A^{0}$ and the charged Higgs boson $H^{\pm}$,
with the color palette representing the new CP-even scalar mass $m_{\eta}$
ranging from $50$ to $400\,~\textrm{~\textrm{GeV}}$. The mass of the CP-odd
scalar spans a wide range from approximately $400<m_{A^{0}}<600\,~\textrm{~\textrm{GeV}}$,
while $m_{H^{\pm}}$ is in the range $140<m_{H^{\pm}}<600\,~\textrm{~\textrm{GeV}}$.
Where the upper branch displays a clear positive correlation between
$m_{A^0}$ and $m_{H^{\pm}}$, extending up to $m_{A^0}\approx~580~\textrm{GeV}$
and $m_{H^{\pm}}\approx~600~\textrm{GeV}$. Along this linear relation,
the mass scale of the new CP-even scalar $m_{\eta}$ gradually shifts
from low values $50$ to $125\,~\textrm{~\textrm{GeV}}$ toward heavier states
($m_{\eta}\sim~250-350~\textrm{GeV}$). Conversely, the lower branch
features an inverse trend where increasing $m_{A^0}$ yields a sharp
decrease in $m_{H^{\pm}}$, terminating in a localized cusp around
$m_{A^0}\approx~545~\textrm{GeV}$ and $m_{H^{\pm}}\approx~140~\textrm{GeV}$.
Within this downward-pointing region, $m_{\eta}$ reaches its lower
bound near $125~\textrm{GeV}$, whereas intermediate regions of this
branch permit heavier scalar masses up to $m_{\eta}\approx~400~\textrm{GeV}$.
The exclusion of large regions surrounding these two branches is driven
primarily by the requirement to satisfy the EWPT, particularly the
oblique $T$ and $S$ parameters, which tightly constrain the mass
splittings among the scalar doublets to preserve approximate custodial
symmetry. Furthermore, theoretical consistency conditions such as
vacuum stability and perturbative unitarity alongside direct collider
search bounds from the Large Hadron Collider tightly delineate the
outer boundaries, ensuring that only physical parameter combinations
yielding viable phenomenology survive.

The masses and mixing values shown here result from a combination of theoretical
and experimental constraints, including the counterterm
definitions. However, some constraints can have a direct impact on
certain results. For instance, large values of both $m_{A^{0}},m_{H^{\pm}}>600\,~\textrm{~\textrm{GeV}}$
are excluded mainly by electroweak precision tests and obviously,
smaller values of the CP-odd scalar mass $m_{A^{0}}<m_{h}/2$ are forbidden
due to the bound on the undetermined Higgs branching ratio $B_{und}<0.22$~\cite{CMS:2015chx}.
This splitting highlights the strong influence of the scalar mass
$m_{\eta}$ on the allowed spectrum of the charged Higgs mass $m_{H^{\pm}}$
relative to $m_{A^{0}}$.

For the extra CP-even scalar, both the PRHM ($m_{\eta}>m_{h}$) and
light dilaton ($m_{\eta}<m_{h}$) scenarios are possible, as shown
in the top-left Fig.~\ref{Mass}; however, the latter is excluded
in the range $m_{\eta}<50\,\textrm{~\textrm{GeV}}$, by Higgs measurements,
where the condition in Eq.~(\ref{eq:XSM}) is violated by the corresponding
CP-even scalar mixing $s^{2}_{ev}$ values.

These patterns reflect the constrained parameter space of the model,
demonstrating how the scalar potential and precision measurement bounds
shape the mass spectrum. The observed hierarchy and correlations have
significant implications for both the theoretical consistency of the
\textit{SI2HDM} and its experimental signatures at colliders.

In the lower left panel of Fig.~\ref{Mass}, we show the mixing parameter
$t_{\beta}$ versus the charged Higgs mass $m_{H^{\pm}}$ (in~\textrm{GeV}),
with the color palette indicating the mixing in the charged sector
($s^{2}_{ch}$). This panel clearly shows that all BPs in the
model satisfy the experimental constraints from flavor physics, specifically
the bounds from the decay $B^{0}_{d}\to\mu^{+}\mu^{-}$~\cite{Enomoto:2015wbn}.
This indicates that all BPs satisfy this constraint, lying in the
region above the red dashed line.

For these BPs, the charged-Higgs mass $m_{H^{\pm}}$ ranges from $140$
to $600$~\textrm{GeV}, and the tree-level mixing parameter $t_{\beta}$ is
between $1.22$ and $4.04$, corresponding to a range of the squared
charged mixing $s^{2}_{ch}$ from $0.60$ to $0.95$. Therefore, the
flavor constraints exclude or disfavor small values of $t_{\beta}$
(below $\sim~$$1.22$), suggesting that the permitted parameter space
is limited by the $B^{0}_{d}\to\mu^{+}\mu^{-}$ measurements. The
lower right panel of Fig.~\ref{Mass} shows a clear anticorrelation
between the squared CP-even mixing ($s^{2}_{ev}$) and the squared
CP-odd mixing ($s^{2}_{od}$). The color palette indicates that the
squared mixing in the charged sector ($s^{2}_{ch}$) is compatible
with the squared CP-odd mixing ($s^{2}_{od}$).

\begin{figure}[t]
\includegraphics[width=0.5\textwidth]{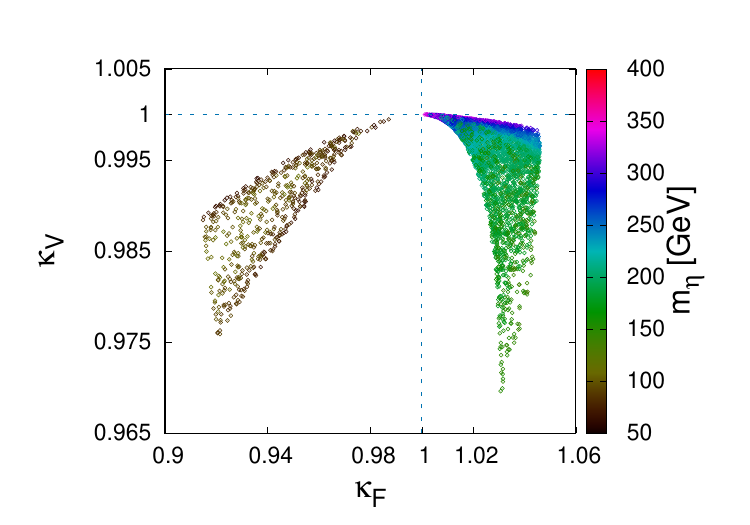}~\includegraphics[width=0.5\textwidth]{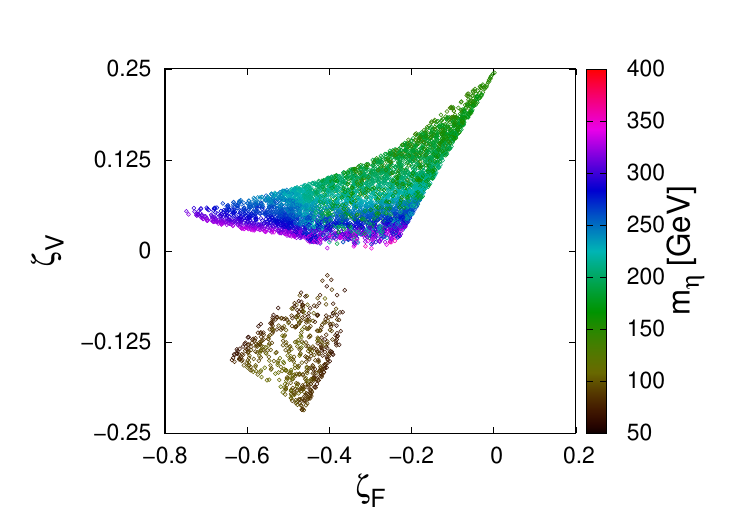}
\caption{The normalized coupling strengths of the Higgs boson (left) to the
gauge boson pairs $\kappa_{V}$ and the fermions $\kappa_{F}$; and
the normalized coupling strengths of the new CP-even scalar (right)
to gauge boson pairs $\zeta_{V}$ and fermions $\zeta_{F}$, with
the color palette indicating the additional CP-even scalar mass $m_{\eta}$
(in~\textrm{GeV}).}
\label{kappa} 
\end{figure}

Fig.~\ref{kappa} presents the normalized coupling strengths of the
Higgs boson (left panel) and the new CP-even scalar (right panel)
to gauge boson pairs and fermions, with the color palette representing
the new CP-even scalar mass $m_{\eta}$ in~\textrm{GeV}.

The Higgs coupling modifier $\kappa_{V}$ is highly constrained, where
in its deviation from the SM is at most $3\%$, while the deviation
of $\kappa_{F}$ from the SM can reach up to $4.6\%$ ($8.6\%$) in
the PRHM (light dilaton) scenario, respectively. These deviations
in $\kappa_{V}$ and $\kappa_{F}$ relative to the SM are driven by
experimental constraints, particularly from the diphoton Higgs decay,
limits on the total Higgs decay width, the Higgs signal strength modifications,
and the bounds derived from the chi-squared $\chi^{2}$ function.
The \textit{SI2HDM} is similar to most SM scalar extensions in which
the couplings of the new CP-even scalar $\eta$ to the fermions and
the gauge bosons are significantly smaller than the SM values. Specifically,
$\zeta_{F}$ is negative within the range $-0.75\leq\zeta_{F}\leq0$,
while $\zeta_{V}$ has two isolated ranges $-0.2\leq\zeta_{V}\leq-0.03$
and $0\leq\zeta_{V}\leq0.24$. As illustrated in Fig.~\ref{kappa}-right,
negative (positive) values of $\kappa_{F}$ correspond to the PRHM
(light dilaton) scenario. This ensures the agreement with all null
results from searches for a heavy resonance.

\begin{figure}[h]
\centering \includegraphics[width=0.5\textwidth]{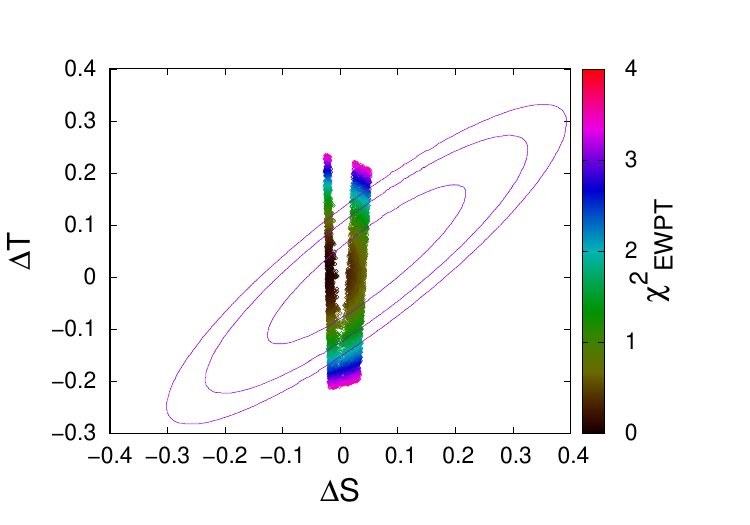}~\includegraphics[width=0.5\textwidth]{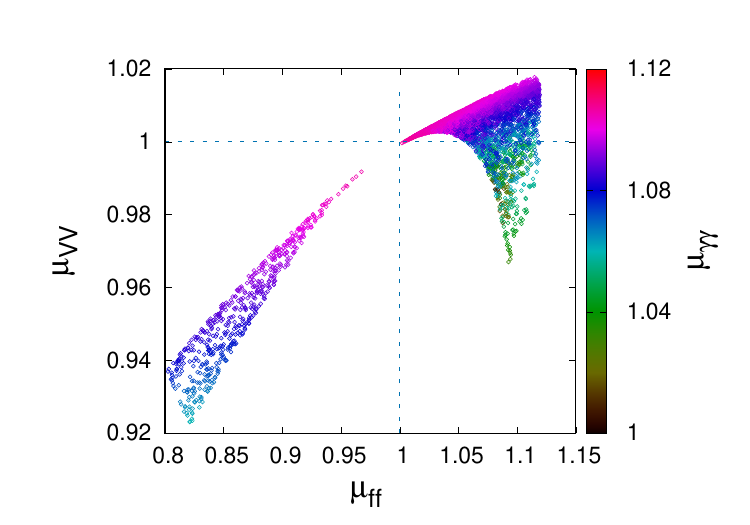}\caption{\textbf{Left panel}: The oblique parameters $\Delta T$ and $\Delta S$
together with the $68\%$, $95\%$ and $99\%$ confidence level ellipses
from a global electroweak fit. The color palette shows the chi-squared
function $\chi^{2}_{\text{EWPT}}$. \textbf{Right panel}: The Higgs
signal strengths $\mu_{VV}$ versus $\mu_{ff}$, with the color palette
representing the signal strength for the diphoton channel $\mu_{\gamma\gamma}$.}
\label{obl}
\end{figure}

In Fig.~\ref{obl}, we present the oblique parameters (left) and
the Higgs modifiers strength (right).

Fig.~\ref{obl}-left illustrates the constraints imposed by the electroweak
precision observables on the parameter space of the \textit{SI2HDM}
through the oblique parameters ($\Delta S$, $\Delta T$), and the
chi-squared function $\chi^{2}_{\text{EWPT}}$. The plot displays
5k BPs, together with the $68\%$, $95\%$, and $99\%$ CL ellipses
derived from global EWPT. The distribution of BPs demonstrates a good
consistency with the electroweak constraints, and the majority of points
lie within the $95\%$ CL region. The vertical spread arises from
the sensitivity of the $T$ parameter to mass splittings between the
charged ($H^{\pm}$) and neutral scalars. These splittings incorporate
full one-loop corrections to scalar masses, enhancing the reliability
of the results. The values of $\Delta S$ are tightly clustered around
positive values near zero, as they depend only weakly on mass splittings,
primarily through logarithmic terms. The accompanying color palette,
representing the $\chi^{2}_{\text{EWPT}}$ function, further corroborates
this conclusion, showing that most BPs exhibit relatively low $\chi^{2}_{\text{EWPT}}$
values, corresponding to good agreement with experimental data. Here,
the values of $\Delta U$ are minimal, with a range of $-0.004\leq\Delta U\leq0.005$.
This indicates that a significant portion of the \textit{SI2HDM} parameter
space remains compatible with current precision measurements.

Fig.~\ref{obl}-right shows the correlation between the vector boson
($\mu_{VV}$) and fermion ($\mu_{ff}$) signal strength modifiers,
with the color palette indicating the corresponding diphoton channel
signal strength $\mu_{\gamma\gamma}$. As illustrated, while current
LHC data~\cite{ParticleDataGroup:2024cfk} constrain the values of
$\mu_{\gamma\gamma}$ across the parameter space, the signal strength
$\mu_{VV}$ exhibits deviations from SM predictions. These deviations
lie between a $7.7\,\%$ suppression and a $1.8\,\%$ enhancement
relative to the SM prediction. Furthermore, the signal strength $\mu_{ff}$
demonstrates a more substantial enhancement, reaching up to $12\,\%$
and the suppression reaching up to $20\,\%$ beyond the SM expectation.
These modifications stem from two sources: experimental bounds on
Higgs signal strengths and the effects of CP-even scalar mixing.

\begin{figure}[t]
\includegraphics[width=0.33\textwidth]{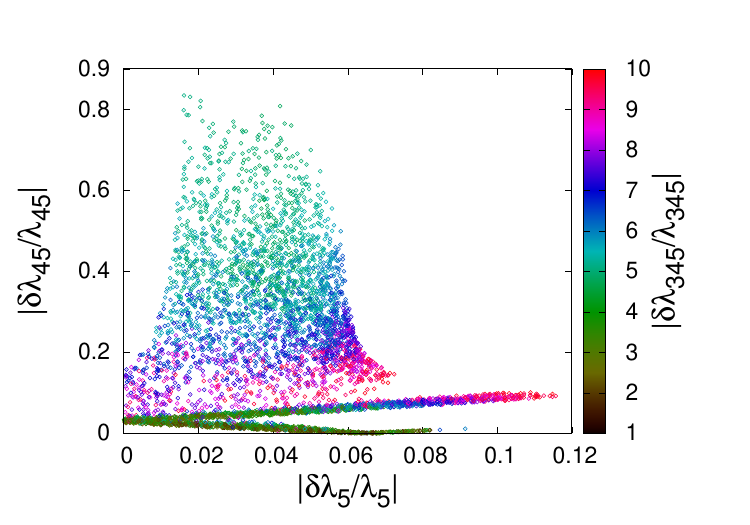}~\includegraphics[width=0.33\textwidth]{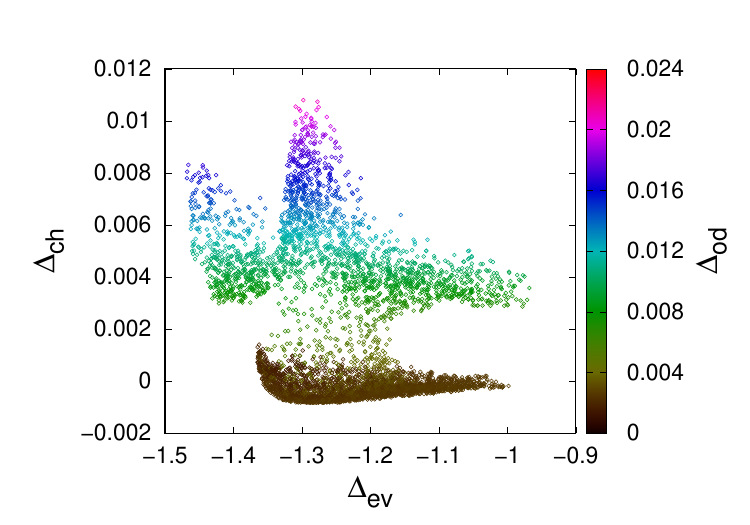}~\includegraphics[width=0.33\textwidth]{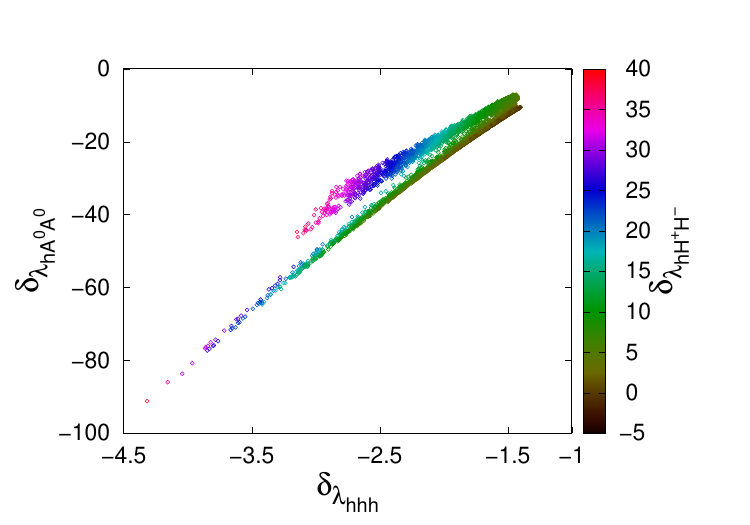}\caption{\textbf{Left panel}: The counterterms $\left|\delta\lambda_{45}/\lambda_{45}\right|$
versus $\left|\delta\lambda_{5}/\lambda_{5}\right|$, with the palette
showing $\left|\delta\lambda_{345}/\lambda_{345}\right|$. \textbf{Middle
panel}: The differences in mixing angles $\Delta_{ev}=s_{ev}-s_{\beta}$,
$\Delta_{ch}=s_{ch}-s_{\beta}$, and $\Delta_{od}=s_{od}-s_{\beta}$
for each sector. \textbf{Right panel}: The relative coupling differences
$\delta_{\lambda_{hXX}}=(\lambda^{1-\ell}_{hXX}-\lambda^{\text{tree}}_{hXX})/\lambda^{\text{tree}}_{hXX}$
for $X\equiv h,A^{0},H^{\pm}$.}
\label{angles}
\end{figure}

In Fig.~\ref{angles}, we present the radiative corrections contribution
to different observables.

Fig.~\ref{angles}-left shows the correlation between the normalized
absolute variations of the couplings $\left|\delta\lambda_{45}/\lambda_{45}\right|$
and $\left|\delta\lambda_{5}/\lambda_{5}\right|$, with the color
palette representing $\left|\delta\lambda_{345}/\lambda_{345}\right|$.
A notable feature is that the allowed BPs are strictly bounded within
$\left|\delta\lambda_{45}/\lambda_{45}\right|<0.85$ and $\left|\delta\lambda_{5}/\lambda_{5}\right|<0.12$,
so the smallness of these variations indicates that the counterterms
$\delta\lambda_{45}$ and $\delta\lambda_{5}$, and consequently the
charged Higgs mass $m_{H^{\pm}}$ and CP-odd scalar mass $m_{A^{0}}$,
remains remarkably stable under one-loop corrections. This reflects
a strong radiative stability, likely arising from theoretical constraints
or underlying symmetries in the scalar potential (see Eq.~(\ref{eq:V1l})).
In contrast, the values of $\left|\delta\lambda_{345}/\lambda_{345}\right|$
span a wider range, ranging from $1$ to $10$. This indicates that
the mass spectrum of the new scalar $m_{\eta}$ is more susceptible
to loop level modifications, with important implications for phenomenological
predictions and experimental bounds.

Fig.~\ref{angles}-middle displays the deviations between the one-loop
corrected mixing angles $s_{Q}$ and their tree-level scalar mixing
$s_{\beta}$. The analysis demonstrates that the mixing angles in
the CP-odd and charged sectors, $s_{od}$ and $s_{ch}$ preserve their
tree-level value $s_{\beta}$ at the one-loop level in certain
regions of the parameter space. In contrast, they deviate significantly
($\Delta_{od}\sim~0.024$, $\Delta_{ch}\sim~0.011$) in regions where
the CP-even sector deviation reaches ( $\Delta_{ev}\sim~-1.25$). Fig.~\ref{angles}-right
presents the correlation between the relative coupling differences
$\delta_{\lambda_{hhh}}=(\lambda^{1-\ell}_{hhh}-\lambda^{\text{tree}}_{hhh})/\lambda^{\text{tree}}_{hhh}$,
$\delta_{\lambda_{hAA}}=(\lambda^{1-\ell}_{hAA}-\lambda^{\text{tree}}_{hAA})/\lambda^{\text{tree}}_{hAA}$,
and $\delta_{\lambda_{hH^{\pm}H^{\pm}}}=(\lambda^{1-\ell}_{hH^{\pm}H^{\pm}}-\lambda^{\text{tree}}_{hH^{\pm}H^{\pm}})/\lambda^{\text{tree}}_{hH^{\pm}H^{\pm}}$.
As the plot shows, both the $\delta_{\lambda_{hhh}}$ and $\delta_{\lambda_{hAA}}$
relative deviations are negative, confirming that one-loop radiative
corrections reduce the trilinear couplings relative to their tree-level
values. In contrast, $\delta_{\lambda_{hH^{\pm}H^{\pm}}}$ spans both
negative and positive values, demonstrating that quantum corrections
can either suppress or enhance the Higgs-charged Higgs trilinear coupling
($hH^{\pm}H^{\pm}$) depending on the region of parameter space. For
example, the enhancement can be as large as $4000\,\%$ for the $hH^{\pm}H^{\pm}$
coupling, whereas the reduction can be as large as $500\,\%$ for
the same coupling, $9200\,\%$ for the $hA^{0}A^{0}$ coupling, and
$440\,\%$ for the $hhh$ coupling. This behavior is consistent with
the expectations of the PRHM scenario, where radiative effects substantially
alter the scalar potential structure at the quantum level~\cite{Ahriche:2021frb}.
The future experimental searches at high-energy colliders can focus
on these significant departures from SM expectations, where accurate
measurements of trilinear Higgs couplings could serve as indirect
probes of BSM physics.

\begin{figure}[h]
\centering \includegraphics[width=0.5\textwidth]{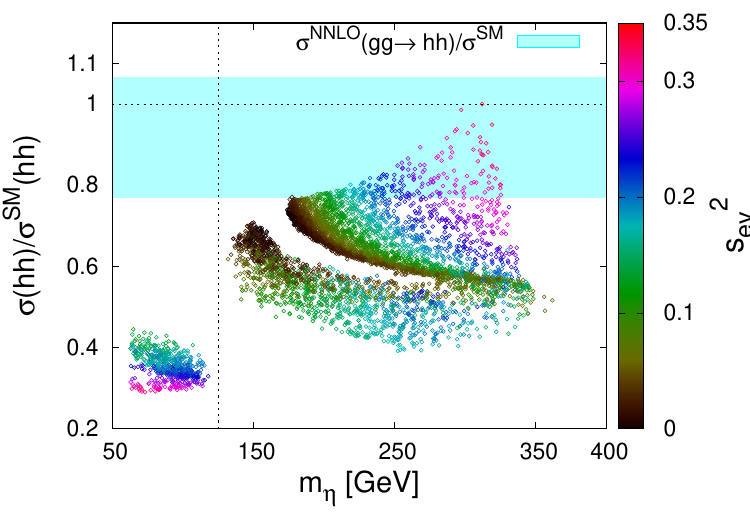}~\includegraphics[width=0.5\textwidth]{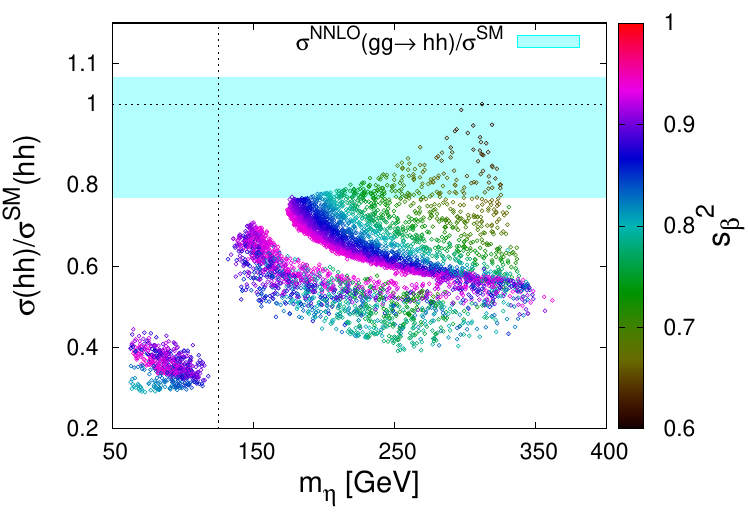}\caption{The di-Higgs production cross section via gluon-gluon fusion at the
LHC with $\sqrt{s}=13\,\mathrm{TeV}$, scaled by the SM value, as
a function of the new CP-even scalar mass $m_{\eta}$ (in~\textrm{GeV}), with
the color palette representing the CP-even mixing parameter $s_{ev}$
(left) and the tree-level mixing parameter $s_{\beta}$. The light
blue band with the black dashed line represents the allowed range
of the ratio $\sigma_{ggF}(hh)/\sigma^{SM}_{ggF}(hh)$ by the recent
measurements~\cite{deFlorian:2019app}.}
\label{dihi} 
\end{figure}

In Fig.~\ref{dihi}, we present the di-Higgs production cross section
via gluon-gluon fusion at the LHC with $\sqrt{s}=13\,\mathrm{TeV}$,
scaled by the SM values as a function of the new CP-even scalar mass,
where the palette shows the CP-even in the left (tree-level in the
right) panel. 

From Fig.~\ref{dihi}, the di-Higgs production at the 13~TeV LHC shows no
significant enhancement for both light dilaton and PRHM scenarios.
The parameter space exhibits two distinct regions: the first concerns
the dilaton scenario ($m_{\eta}<m_{h}$) with the cross section reduction
$0.3<\sigma_{ggF}/\sigma^{SM}_{ggF}<0.45$. The second, more extended
region concerns the PRHM scenario ($m_{h}<m_{\eta}$) with cross-section
reduction $0.4<\sigma_{ggF}/\sigma^{SM}_{ggF}<1$. In the first region
that is, in the dilaton scenario, the suppression ranges
from $55\%$ to $70\%$, with the squared mixing angle in the CP-even
sector $s^{2}_{ev}$ is in the range $0.12$ and $0.33$. For the
PRHM scenario, the di-Higgs production cross section can be suppressed by up to
$60\%$ as the mixing angle in the CP-even sector becomes larger.
This reduction arises from the Higgs ($\eta$) coupling modifier to
fermion pairs, $\kappa_{F}$ ($\zeta_{F}$), and the Higgs trilinear
self couplings $\lambda_{hhh}$. Notably, not all BPs would be excluded
by current experimental data~\cite{deFlorian:2019app}. If di-Higgs production is observed with the SM cross section or a larger
one, the light-dilaton scenario would be fully excluded. However, if
a reduced cross section is observed, our setup would have a better
chance of being probed.

To probe this model at colliders, the production and decays of the
BSM particles ($\eta$, $H^{\pm}$, and $A^{0}$) in pp collisions
must be investigated. At the LHC, a light charged Higgs boson ($m_{H^{\pm}}<600~\textrm{~\textrm{GeV}}$)
would be produced mainly via top quark decays: $pp\to t\bar{t}$,
with $t\to bH^{+}$ (provided $m_{H^{\pm}}<m_{t}-m_{b}$). For the
CP-odd scalar $A^{0}$ across a wide mass range ($400~\textrm{~\textrm{GeV}}\sim~600~\textrm{~\textrm{GeV}}$),
the leading production mechanism is gluon fusion ($gg\to A^{0}$).
This loop-induced process has a cross section proportional to $(\frac{c_{\beta_{ev}}}{s_{\beta}})^{2}$
and is dominant for most of the mass range due to the high gluon luminosity.
For the heavy CP-even scalar $m_{\eta}>m_{h}~\textrm{~\textrm{GeV}}$, the main
production channels are ggF ($gg\to\eta$) and VBF ($qq\to jj\eta$).
While ggF typically has the largest cross section, VBF provides a
distinct experimental signature with two forward jets. These signals
depend strongly on the model's parameters (masses, mixing angles,
and $t_{\beta}$). A detailed investigation is therefore needed to
explore them across the viable parameter space.

These results address the fundamental question raised in the introduction:
whether an additional real singlet scalar is necessary to achieve
radiatively driven EWSB in SI models. Our comprehensive analysis of
the \textit{SI2HDM} demonstrates that electroweak symmetry breaking
and a consistent scalar mass spectrum can be realized without introducing
an additional singlet scalar. In particular, the model accommodates
both the light-dilaton and PRHM scenarios within its viable parameter
space, with the observed Higgs boson corresponding to the heavier
and lighter CP-even states, respectively. The model's viability under
stringent theoretical and experimental constraints, together with
its distinctive predictions for the trilinear Higgs couplings and
di-Higgs production, demonstrates that an additional singlet is not
a mandatory ingredient. This establishes the \textit{SI2HDM} as a
minimal and self-contained framework for addressing the hierarchy
problem through classical scale invariance.

\section{Conclusion~\label{sec:Conc}}

In this work, we have presented a comprehensive analysis of the Scale Invariant Two-Higgs-Doublet Model (\textit{SI2HDM}), where electroweak symmetry breaking is radiatively induced without introducing an additional real singlet scalar. After deriving the one-loop effective potential and the corresponding corrected mass matrices, we systematically applied the full set of theoretical and experimental constraints, including perturbative unitarity, vacuum stability, electroweak precision tests, Higgs decay rates, signal strength modifiers, and experimental flavor constraints, especially the bounds on the decay $B^{0}_{d}\to\mu^{+}\mu^{-}$.

Our investigation reveals that the viable parameter space of the \textit{SI2HDM} is highly constrained and phenomenologically distinct from the conventional 2HDM. A key finding is the existence of two distinct scenarios: the light-dilaton scenario ($50~\textrm{GeV} \le m_\eta < m_h$), in which the new CP-even scalar is lighter than the observed SM-like Higgs boson, and the PRHM scenario ($m_\eta > m_h$), in which the observed Higgs boson corresponds to the lighter CP-even state and acquires its mass radiatively. Within this viable space, the charged Higgs mass is tightly bounded to $140 \le m_{H^\pm} \le 600~\textrm{GeV}$, the CP-odd scalar mass spans $400 \le m_{A^0} \le 600~\textrm{GeV}$, and the tree-level VEV ratio is restricted to $1.22 \le \tan\beta \le 4.04$. We demonstrated a crucial difference in radiative stability: the charged Higgs mass exhibits limited sensitivity to loop corrections ($\left|\delta\lambda_{45}/\lambda_{45}\right|<0.85$), while the CP-odd scalar mass is significantly more susceptible, governed primarily by the counterterm $\delta\lambda_{5}$.

The model's phenomenology is marked by distinctive signatures. The analysis of electroweak precision observables shows good agreement with data, with the majority of benchmark points falling within the $95\%$ confidence level, as the structured mass spectrum keeps unwanted contributions to the $T$-parameter under control. Furthermore, one-loop radiative corrections suppress the trilinear couplings $\lambda_{hhh}$ and $\lambda_{hA^{0}A^{0}}$, while the $\lambda_{hH^{\pm}H^{\pm}}$ coupling can be either enhanced (up to $4000\%$) or suppressed depending on the parameter space. These corrections, combined with modifications to the Higgs fermion couplings, lead to a significant reduction in the di-Higgs production cross section at the 13~TeV LHC, by $55\%\text{--}70\%$ in the light-dilaton scenario and up to $60\%$ in the PRHM scenario compared to the SM prediction, without being excluded by current data.

These results underscore the predictive power and testability of the \textit{SI2HDM} framework. Crucially, this work demonstrates that a singlet scalar is not a necessary ingredient for a viable, radiatively driven EWSB mechanism. The \textit{SI2HDM} provides a minimal and fully functional alternative, with a tightly constrained parameter space compared to models assisted by a real singlet. Unlike singlet-extended models, the \textit{SI2HDM} accommodates both the light-dilaton and PRHM scenarios within its viable parameter space. This constrained structure enhances the model's predictability, leading to concrete phenomenological predictions, such as the substantial suppression of di-Higgs production. Consequently, the model offers a well-defined and distinctive framework for radiatively driven electroweak symmetry breaking, making it testable and distinguishable at the LHC and future colliders through precise measurements of Higgs pair production and the properties of the extended scalar sector.

\subsection*{Acknowledgments}

The work of N.B. is supported by the Algerian Ministry of Higher Education
and Scientific Research under the PRFU Project No. \textit{B00L02UN180120230001}.
A.A. is funded by the University of Sharjah via the HEP RG operational grant.

\appendix

\section{The one-loop Parameters~\label{sec:Mass}}

The one-loop scalar potential can be written in function of the CP-even,
CP-odd scalar and charged scalar fields as in Eq.~(\ref{eq:V1l}).
By using the one-loop tadpole condition $\partial V^{1-\ell}(\phi_1 ,\phi_2 )/\partial\phi_1 =\partial V^{1-\ell}(\phi_1 ,\phi_2 )/\partial\phi_2 =0$
at the vacuum ($<\phi_i >=\upsilon_i $ and $<a_i >=<\phi^{\pm}_i >=0$),
one writes the counterterms $\delta\lambda_1 $ and $\delta\lambda_2 $
in function of $\delta\lambda_{345}$; and a combination of the field
dependent masses, $m^{2}_i (\Phi_{1,2})$, and their derivatives
as 
\begin{align}
\delta\lambda_1 = & -3\delta\lambda_{345}t^{2}_{\beta}-\frac{3}{16\pi^{2}\upsilon^{3}c^{3}_{\beta}}\sum_i n_i m^{2}_i \big(m^{2}_i \big)_{,\phi_1}\Big[\log(m^{2}_i /\Lambda^{2})-1\Big],\nonumber \\
\delta\lambda_2 = & -3\delta\lambda_{345}/t^{2}_{\beta}-\frac{3}{16\pi^{2}\upsilon^{3}s^{3}_{\beta}}\sum_i n_i m^{2}_i \big(m^{2}_i \big)_{,\phi_2}\Big[\log(m^{2}_i /\Lambda^{2})-1\Big],
\end{align}
with $Q_{,x}=\frac{\partial Q}{\partial x}$. Then, one writes
\begin{align}
\delta\lambda_1 = & -3\delta\lambda_{345}t^{2}_{\beta}-\frac{3}{16\pi^{2}\upsilon^{2}c^{3}_{\beta}}\sum_i n_i m^{2}_i \left[\alpha_i c_{b}+\sigma_i s_{b}\right]\Big[\log(m^{2}_i /\Lambda^{2})-1\Big],\nonumber \\
\delta\lambda_2 = & -3\delta\lambda_{345}/t^{2}_{\beta}-\frac{3}{16\pi^{2}\upsilon^{2}s^{3}_{\beta}}\sum_i n_i m^{2}_i \left[\beta_i s_{b}+\sigma_i c_{b}\right]\Big[\log(m^{2}_i /\Lambda^{2})-1\Big],
\end{align}

In addition, the parameters $a_{Q},~b_{Q}$ and $c_{Q}$ used in~\ref{M1-1}
are defined by 
\begin{align}
a_{0^{+}} & =\frac{1}{32m^{2}_{h}\pi^{2}}\sum_i n_i \left\{m^{2}_i \left(\big(m^{2}_i \big)_{,\phi_1 ,\phi_1}-\frac{3}{\upsilon c_{\beta}}\big(m^{2}_i \big)_{,\phi_1}\right)\Big[\log(m^{2}_i /\Lambda^{2})-1\Big]+\big(m^{2}_i \big)^{2}_{,\phi_1}\log(m^{2}_i /\Lambda^{2})\right\} ,\nonumber \\
b_{0^{+}} & =\frac{1}{32m^{2}_{h}\pi^{2}}\sum_i n_i \left\{m^{2}_i \left(\big(m^{2}_i \big)_{,\phi_2 ,\phi_2}-\frac{3}{\upsilon s_{\beta}}\big(m^{2}_i \big)_{,\phi_2}\right)\Big[\log(m^{2}_i /\Lambda^{2})-1\Big]+\big(m^{2}_i \big)^{2}_{,\phi_2}\log(m^{2}_i /\Lambda^{2})\right\} ,\nonumber \\
c_{0^{+}} & =\frac{1}{32m^{2}_{h}\pi^{2}}\sum_i n_i \left\{\left(\big(m^{2}_i \big)_{,\phi_1}\big(m^{2}_i \big)_{,\phi_2}+m^{2}_i \big(m^{2}_i \big)_{,\phi_1 ,\phi_2}\right)\log(m^{2}_i /\Lambda^{2})-m^{2}_i \big(m^{2}_i \big)_{,\phi_1 ,\phi_2}\right\} .\label{eq:ap}
\end{align}
\begin{align}
a_{0^{-}} & =\frac{1}{32\pi^{2}m^{2}_{A^{0}}}\sum_i n_i m^{2}_i \left(\big(m^{2}_i \big)_{,a_1 ,a_1}-\frac{1}{\upsilon c_{\beta}}\big(m^{2}_i \big)_{,\phi_1}\right)\Big[\log(m^{2}_i /\Lambda^{2})-1\Big],\nonumber \\
b_{0^{-}} & =\frac{1}{32\pi^{2}m^{2}_{A^{0}}}\sum_i n_i m^{2}_i \left(\big(m^{2}_i \big)_{,a_2 ,a_2}-\frac{1}{\upsilon s_{\beta}}\big(m^{2}_i \big)_{,\phi_2}\right)\Big[\log(m^{2}_i /\Lambda^{2})-1\Big],\nonumber \\
c_{0^{-}} & =\frac{1}{32\pi^{2}m^{2}_{A^{0}}}\sum_i n_i m^{2}_i \big(m^{2}_i \big)_{,a_1 ,a_2}\Big[\log(m^{2}_i /\Lambda^{2})-1\Big],\label{eq:am}
\end{align}
and 
\begin{align}
a_{\pm} & =\frac{1}{32\pi^{2}m^{2}_{H^{\pm}}}\sum_i n_i m^{2}_i \left(\big(m^{2}_i \big)_{,\phi^{+}_1 ,\phi^{-}_1}-\frac{1}{\upsilon c_{\beta}}\big(m^{2}_i \big)_{,\phi_1}\right)\Big[\log(m^{2}_i /\Lambda^{2})-1\Big],\nonumber \\
b_{\pm} & =\frac{1}{32\pi^{2}m^{2}_{H^{\pm}}}\sum_i n_i m^{2}_i \left(\big(m^{2}_i \big)_{,\phi^{+}_2 ,\phi^{-}_2}-\frac{1}{\upsilon s_{\beta}}\big(m^{2}_i \big)_{,\phi_2}\right)\Big[\log(m^{2}_i /\Lambda^{2})-1\Big],\nonumber \\
c_{\pm} & =\frac{1}{32\pi^{2}m^{2}_{H^{\pm}}}\sum_i n_i m^{2}_i \big(m^{2}_i \big)_{,\phi^{+}_1 ,\phi^{-}_2}\Big[\log(m^{2}_i /\Lambda^{2})-1\Big],\label{eq:apm}
\end{align}
where all formulas Eqs.~(\ref{eq:ap},\ref{eq:am} and \ref{eq:apm})
are estimated at the vacuum $<\phi_i >=\upsilon_i $ and $<a_i >=<\phi^{\pm}_i >=0$.

The trilinear couplings $hA^{0}A^{0}$ and $hH^{+}H^{-}$ can be obtained
at tree-level but using one-loop mixing angles as 
\begin{align}
\frac{\lambda_{hA^{0}A^{0}}}{\upsilon} & =c^{2}_{\beta_{od}}\Big(\frac{1}{3}s_{\beta}c_{\beta_{ev}}\lambda_2 -c_{\beta}s_{\beta_{ev}}(\lambda_{3}+\lambda_{4}-\lambda_{5})\Big)+s^{2}_{\beta_{od}}\Big(s_{\beta}c_{\beta_{ev}}(\lambda_{3}+\lambda_{4}-\lambda_{5})-\frac{1}{3}c_{\beta}s_{\beta_{ev}}\lambda_1 \Big)\nonumber \\
 & -2s_{\beta_{od}}c_{\beta_{od}}(c_{\beta}c_{\beta_{ev}}-s_{\beta}s_{\beta_{ev}})\lambda_{5}\nonumber \\
\frac{\lambda_{hH^{+}H^{-}}}{\upsilon} & =-c^{2}_{\beta_{ch}}\Big(c_{\beta}s_{\beta_{ev}}\lambda_{3}-\frac{1}{3}s_{\beta}c_{\beta_{ev}}\lambda_2 \Big)+s^{2}_{\beta_{ch}}\Big(s_{\beta}c_{\beta_{ev}}\lambda_{3}-\frac{1}{3}c_{\beta}s_{\beta_{ev}}\lambda_1 \Big)\nonumber \\
 & -s_{\beta_{ch}}c_{\beta_{ch}}(c_{\beta}c_{\beta_{ev}}-s_{\beta}s_{\beta_{ev}})\lambda_{45}.\label{eq:LLL}
\end{align}

The one-loop trilinear scalar couplings can be estimated as the $3^{rd}$
derivatives of the effective potential Eq.~(\ref{eq:V1l}) estimated
at the vacuum $<\phi_i >=\upsilon_i $ and $<a_i >=<\phi^{\pm}_i >=0$.
Then, ones writes 
\begin{align}
\lambda_{hhh} & =-s^{3}_{\beta_{ev}}V^{1-\ell}_{,\phi_1 ,\phi_1 ,\phi_1}+3s^{2}_{\beta_{ev}}c_{\beta_{ev}}V^{1-\ell}_{,\phi_1 ,\phi_1 ,\phi_2}-3s_{\beta_{ev}}c^{2}_{\beta_{ev}}V^{1-\ell}_{,\phi_1 ,\phi_2 ,\phi_2}+c^{3}_{\beta_{ev}}V^{1-\ell}_{,\phi_2 ,\phi_2 ,\phi_2}\nonumber \\
\lambda_{hh\eta} & =s^{2}_{\beta_{ev}}c_{\beta_{ev}}V^{1-\ell}_{,\phi_1 ,\phi_1 ,\phi_1}+s_{\beta_{ev}}\Big(1-3c^{2}_{\beta_{ev}}\Big)V^{1-\ell}_{,\phi_1 ,\phi_1 ,\phi_2}+c_{\beta_{ev}}\Big(1-3s^{2}_{\beta_{ev}}\Big)V^{1-\ell}_{,\phi_1 ,\phi_2 ,\phi_2}+c^{2}_{\beta_{ev}}s_{\beta_{ev}}V^{1-\ell}_{,\phi_2 ,\phi_2 ,\phi_2}\nonumber \\
\lambda_{h\eta\eta} & =-s_{\beta_{ev}}c^{2}_{\beta_{ev}}V^{1-\ell}_{,\phi_1 ,\phi_1 ,\phi_1}+c_{\beta_{ev}}\Big(1-3s^{2}_{\beta_{ev}}\Big)V^{1-\ell}_{,\phi_1 ,\phi_1 ,\phi_2}-s_{\beta_{ev}}\Big(1-3c^{2}_{\beta_{ev}}\Big)V^{1-\ell}_{,\phi_1 ,\phi_2 ,\phi_2}+c_{\beta_{ev}}s^{2}_{\beta_{ev}}V^{1-\ell}_{,\phi_2 ,\phi_2 ,\phi_2}\nonumber \\
\lambda_{\eta\eta\eta} & =c^{3}_{\beta_{ev}}V^{1-\ell}_{,\phi_1 ,\phi_1 ,\phi_1}+3s_{\beta_{ev}}c^{2}_{\beta_{ev}}V^{1-\ell}_{,\phi_1 ,\phi_1 ,\phi_2}+3s^{2}_{\beta_{ev}}c_{\beta_{ev}}V^{1-\ell}_{,\phi_1 ,\phi_2 ,\phi_2}+s^{3}_{\beta_{ev}}V^{1-\ell}_{,\phi_2 ,\phi_2 ,\phi_2}.
\end{align}

\section{CP-Odd \& Charged One-Loop Masses~\label{sec:Gold}}

It has been argued in the literature~\cite{Lee:2012jn,Lane:2018ycs,Lane:2019dbc,Eichten:2022vys},
that the $2\times2$ squared mass matrix of both CP-odd and charged
sector has the same value at both tree- and one-loop-levels, by just
fixing a single counterterm\footnote{Actually, in~\cite{Lee:2012jn,Lane:2018ycs,Lane:2019dbc,Eichten:2022vys},
the authors did use explicitly the counterterms notation, but instead
the quartic coupling values were re-estimated after imposing such
renormalization conditions.}. Here, we want to show that this is not possible if the contributions
of the SM-like Higgs, CP-odd scalar and charged scalar ($h,A^{0},H^{\pm}$)
are considered when estimating the one-loop effective potential. In
our work, we took these exact values of the field dependent masses,
i.e., the large value after diagonalizing the $2\times2$ squared
mass matrix. In order to make this point clear, we assume the field
dependent masses have the general form $m^{2}_i =\frac{1}{2}(\alpha_i \phi^{2}_1 +\beta_i \phi^{2}_2 +2\sigma_i \phi_1 \phi_2 +\rho^{(1)}_i a^{2}_1 +\rho^{(2)}_i a^{2}_2 +2\rho^{(3)}_i a_1 a_2 +2\kappa^{(1)}_i \phi^{+}_1 \phi^{-}_1 +2\kappa^{(2)}_i \phi^{+}_2 \phi^{-}_2 +\kappa^{(3)}_i (\phi^{+}_1 \phi^{-}_2 +\phi^{-}_1 \phi^{+}_2 ))$,
where $\alpha_i =\rho^{(1)}_i =\kappa^{(1)}_i $, $\beta_i =\rho^{(2)}_i =\kappa^{(2)}_i $
and $\sigma_i =\rho^{(3)}_i =\kappa^{(3)}_i =0$, for all fields
except for $i=h,A^{0},H^{\pm}$.

The coefficients of the field dependent masses of SM-like Higgs $h$,
the CP-odd scalar $A^{0}$ and the charged scalar $H^{\pm}$given
by
\begin{align}
\alpha_{h} & =\left(\lambda_{3}+\lambda_{4}+\lambda_{5}\right)c^{2}_{\beta}+s^{2}_{\beta}\lambda_1 ,\,\,\beta_{h}=\left(\lambda_{3}+\lambda_{4}+\lambda_{5}\right)s^{2}_{\beta}+c^{2}_{\beta}\lambda_2 ,\,\,\sigma_{h}=-2\left(\lambda_{3}+\lambda_{4}+\lambda_{5}\right)c_{\beta}s_{\beta},\nonumber \\
\rho^{(1)}_{h} & =\left(\lambda_{3}+\lambda_{4}-\lambda_{5}\right)c^{2}_{\beta}+\frac{s^{2}_{\beta}\lambda_1}{3},\,\,\rho^{(2)}_{h}=\left(\lambda_{3}+\lambda_{4}-\lambda_{5}\right)s^{2}_{\beta}+\frac{c^{2}_{\beta}\lambda_2}{3},\,\,\rho^{(3)}_{h}=-2\lambda_{5}c_{\beta}s_{\beta},\nonumber \\
\kappa^{(1)}_{h} & =\lambda_{3}c^{2}_{\beta}+\frac{s^{2}_{\beta}\lambda_1}{3},\,\,\kappa^{(2)}_{h}=\lambda_{3}s^{2}_{\beta}+\frac{c^{2}_{\beta}\lambda_2}{3},\,\,\kappa^{(3)}_{h}=-\left(\lambda_{4}+\lambda_{5}\right)c_{\beta}s_{\beta},
\end{align}
\begin{align}
\alpha_{A^{0}} & =\rho^{(1)}_{h},\,\,\beta_{A^{0}}=\rho^{(2)}_{h},\,\,\sigma_{A^{0}}=\rho^{(3)}_{h},\nonumber \\
\rho^{(1)}_{A^{0}} & =\alpha_{h},\,\,\rho^{(2)}_{A^{0}}=\beta_{h},\,\,\rho^{(3)}_{A^{0}}=\sigma_{h},\nonumber \\
\kappa^{(1)}_{A^{0}} & =\kappa^{(1)}_{h},\,\,\kappa^{(2)}_{A^{0}}=\kappa^{(2)}_{h},\,\,\kappa^{(3)}_{A^{0}}=\kappa^{(3)}_{h},
\end{align}
\begin{align}
\alpha_{H^{\pm}} & =\kappa^{(1)}_{h}=\kappa^{(1)}_{A^{0}},\,\,\beta_{H^{\pm}}=\kappa^{(2)}_{h}=\kappa^{(2)}_{A^{0}},\,\,\sigma_{H^{\pm}}=\sigma_{h}=\rho^{(3)}_{A^{0}},\nonumber \\
\rho^{(1)}_{H^{\pm}} & =\alpha_{H^{\pm}},\,\,\rho^{(2)}_{H^{\pm}}=\beta_{H^{\pm}},\,\,\rho^{(3)}_{H^{\pm}}=\sigma_{H^{\pm}},\nonumber \\
\kappa^{(1)}_{H^{\pm}} & =\left(\lambda_{3}+\lambda_{4}\right)c^{2}_{\beta}+\frac{2s^{2}_{\beta}}{3}\lambda_1 ,\,\,\kappa^{(2)}_{H^{\pm}}=\left(\lambda_{3}+\lambda_{4}\right)s^{2}_{\beta}+\frac{2c^{2}_{\beta}}{3}\lambda_2 ,\,\,\kappa^{(3)}_{H^{\pm}}=-\left(\lambda_{3}+\lambda_{4}+2\lambda_{5}\right)c_{\beta}s_{\beta}.
\end{align}

The coefficients of the field dependent masses of the $W^{\pm}$ and
$Z$ gauge bosons are given by
\begin{align}
\alpha_{W^{\pm}} & =\beta_{W^{\pm}}=\rho^{(1)}_{W^{\pm}}=\rho^{(2)}_{W^{\pm}}=\kappa^{(1)}_{W^{\pm}}=\kappa^{(2)}_{W^{\pm}}=\frac{g^{2}_2}{2},\,\,\sigma_{W^{\pm}}=\rho^{(3)}_{W^{\pm}}=\kappa^{(3)}_{W^{\pm}}=0,\nonumber \\
\alpha_{Z} & =\beta_{Z}=\rho^{(1)}_{Z}=\rho^{(2)}_{Z}=\frac{\left(g^{2}_2 +g^{2}_1 \right)}{2},\,\,\sigma_{Z}=\rho^{(3)}_{Z}=\kappa^{(3)}_{Z}=0,\,\,\kappa^{(1)}_{Z}=\kappa^{(2)}_{Z}=\frac{\left(g_2 -g_1 \right)^{2}\left(g_2 +g_1 \right)^{2}}{2\left(g^{2}_2 +g^{2}_1 \right)}.
\end{align}

For fermions, we have
\begin{align}
\beta_{t} & =\rho^{(2)}_{t}=\frac{y^{2}_{t}}{s^{2}_{b}},\,\,\alpha_{t}=\sigma_{t}=\rho^{(1)}_{t}=\rho^{(3)}_{t}=\kappa^{(1)}_{t}=\kappa^{(3)}_{t}=0,\,\,\kappa^{(2)}_{t}=-\frac{y_{t}y_{b}}{s^{2}_{b}},\nonumber \\
\beta_{b} & =\rho^{(2)}_{b}=\frac{y^{2}_{b}}{s^{2}_{b}},\,\,\alpha_{b}=\sigma_{b}=\rho^{(1)}_{b}=\rho^{(3)}_{b}=\kappa^{(2)}_{b}=\kappa^{(3)}_{b}=0,\,\,\kappa^{(1)}_{b}=-\frac{y_{t}y_{b}}{s^{2}_{b}}.
\end{align}

\section{Perturbative Unitarity Matrices~\label{sec:Matrice}}

Here, we present the relevant scattering amplitude matrices and their
eigenvalues for the unitarity analysis. The scalar scattering matrix
of the neutral states is described by a $14\times14$ matrix in the
basis $\{\frac{\phi_1 \phi_1}{\sqrt{2}},\frac{\phi_2 \phi_2}{\sqrt{2}},\frac{a_1 a_1}{\sqrt{2}},\frac{a_2 a_2}{\sqrt{2}},\phi^{+}_1 \phi^{-}_1 ,\phi^{+}_2 \phi^{-}_2 ,\phi_1 \phi_2 ,a_1 a_2 ,\phi^{+}_1 \phi^{-}_2 ,\phi^{+}_2 \phi^{-}_1 ,\phi_1 a_2 ,\phi_2 a_1 ,\phi_1 a_1 ,\phi_2 a_2 \}$.
However, this matrix can be simplified into three submatrices as 
\begin{equation}
\mathcal{M}^{N}_{14\times14}=\left(\begin{array}{ccc}
\left(\mathcal{M}^{N}_1 \right)_{6\times6} & 0 & 0\\
0 & \left(\mathcal{M}^{N}_2 \right)_{6\times6} & 0\\
0 & 0 & \left(\mathcal{M}^{N}_{3}\right)_{2\times2}
\end{array}\right),
\end{equation}
where the matrix in the basis $\left\{\frac{\phi_1 \phi_1}{\sqrt{2}},\frac{\phi_2 \phi_2}{\sqrt{2}},\frac{a_1 a_1}{\sqrt{2}},\frac{a_2 a_2}{\sqrt{2}},\phi^{+}_1 \phi^{-}_1 ,\phi^{+}_2 \phi^{-}_2 \right\} $
is given by

\begin{equation}
\mathcal{M}^{N}_1 =\left(\begin{array}{cccccc}
\frac{\lambda_1}{2} & \frac{\lambda_{3}+\lambda_{4}+\lambda_{5}}{2} & \frac{\lambda_1}{6} & \frac{\lambda_{3}+\lambda_{4}-\lambda_{5}}{2} & \frac{\lambda_1}{3\sqrt{2}} & \frac{\lambda_{3}}{\sqrt{2}}\\
\frac{\lambda_{3}+\lambda_{4}+\lambda_{5}}{2} & \frac{\lambda_2}{2} & \frac{\lambda_{3}+\lambda_{4}-\lambda_{5}}{2} & \frac{\lambda_2}{6} & \frac{\lambda_{3}}{\sqrt{2}} & \frac{\lambda_2}{3\sqrt{2}}\\
\frac{\lambda_1}{6} & \frac{\lambda_{3}+\lambda_{4}-\lambda_{5}}{2} & \frac{\lambda_1}{2} & \frac{\lambda_{3}+\lambda_{4}+\lambda_{5}}{2} & \frac{\lambda_1}{3\sqrt{2}} & \frac{\lambda_{3}}{\sqrt{2}}\\
\frac{\lambda_{3}+\lambda_{4}-\lambda_{5}}{2} & \frac{\lambda_2}{6} & \frac{\lambda_{3}+\lambda_{4}+\lambda_{5}}{2} & \frac{\lambda_2}{2} & \frac{\lambda_{3}}{\sqrt{2}} & \frac{\lambda_2}{3\sqrt{2}}\\
\frac{\lambda_1}{3\sqrt{2}} & \frac{\lambda_{3}}{\sqrt{2}} & \frac{\lambda_1}{3\sqrt{2}} & \frac{\lambda_{3}}{\sqrt{2}} & \frac{2\lambda_1}{3} & \lambda_{3}+\lambda_{4}\\
\frac{\lambda_{3}}{\sqrt{2}} & \frac{\lambda_2}{3\sqrt{2}} & \frac{\lambda_{3}}{\sqrt{2}} & \frac{\lambda_2}{3\sqrt{2}} & \lambda_{3}+\lambda_{4} & \frac{2\lambda_2}{3}
\end{array}\right),
\end{equation}

that can be partially diagonalized to get the eigenvalues $\left(\frac{\lambda_1 +\lambda_2}{3}\right)\pm\sqrt{\left(\frac{\lambda_1 -\lambda_2}{3}\right)^{2}+\left(2\lambda_{5}\right)^{2}}$,
while the remaining ones should be obtained numerically.

The second submatrix $\mathcal{M}^{N}_2 $ that is defined in the
basis $\left\{\phi_1 \phi_2 ,a_1 a_2 ,\phi^{+}_1 \phi^{-}_2 ,\phi^{+}_2 \phi^{-}_1 ,\phi_1 a_2 ,\phi_2 a_1 \right\} $
is given by 
\begin{align}
\mathcal{M}^{N}_2  & =\left(\begin{array}{cccccc}
\lambda_{3}+\lambda_{4}+\lambda_{5} & \lambda_{5} & \frac{\lambda_{4}+\lambda_{5}}{2} & \frac{\lambda_{4}+\lambda_{5}}{2} & 0 & 0\\
\lambda_{5} & \lambda_{3}+\lambda_{4}+\lambda_{5} & \frac{\lambda_{4}+\lambda_{5}}{2} & \frac{\lambda_{4}+\lambda_{5}}{2} & 0 & 0\\
\frac{\lambda_{4}+\lambda_{5}}{2} & \frac{\lambda_{4}+\lambda_{5}}{2} & \lambda_{3}+\lambda_{4} & 2\lambda_{5} & \frac{-i\left(\lambda_{4}-\lambda_{5}\right)}{2} & \frac{i\left(\lambda_{4}-\lambda_{5}\right)}{2}\\
\frac{\lambda_{4}+\lambda_{5}}{2} & \frac{\lambda_{4}+\lambda_{5}}{2} & 2\lambda_{5} & \lambda_{3}+\lambda_{4} & \frac{i\left(\lambda_{4}-\lambda_{5}\right)}{2} & \frac{-i\left(\lambda_{4}-\lambda_{5}\right)}{2}\\
0 & 0 & \frac{i\left(\lambda_{4}-\lambda_{5}\right)}{2} & \frac{-i\left(\lambda_{4}-\lambda_{5}\right)}{2} & \lambda_{3}+\lambda_{4}-\lambda_{5} & \lambda_{5}\\
0 & 0 & \frac{-i\left(\lambda_{4}-\lambda_{5}\right)}{2} & \frac{i\left(\lambda_{4}-\lambda_{5}\right)}{2} & \lambda_{5} & \lambda_{3}+\lambda_{4}-\lambda_{5}
\end{array}\right),
\end{align}
that leads to the eigenvalues $\lambda_{3}+\lambda_{4}$, $\lambda_{3}\pm\lambda_{5}$
and $\lambda_{3}+2\lambda_{4}\pm3\lambda_{5}$. The third submatrix
$\mathcal{M}^{N}_{3}$ corresponds to the basis $\left\{\phi_1 a_1 ,\phi_2 a_2 \right\} $
is given by 
\begin{align}
\mathcal{M}^{N}_{3} & =\left(\begin{array}{cc}
\frac{\lambda_1}{3} & \lambda_{5}\\
\lambda_{5} & \frac{\lambda_2}{3}
\end{array}\right),
\end{align}
which gives $\frac{\lambda_1 +\lambda_2}{6}\pm\sqrt{\left(\frac{\lambda_1 -\lambda_2}{6}\right)^{2}+\lambda^{2}_{5}}$
as eigenvalues.

The scalar scattering matrix of the charged states is described by
a $8\times8$ matrix defined in the basis $\left\{\phi_1 \phi^{+}_1 ,\phi_2 \phi^{+}_2 ,a_1 \phi^{+}_1 ,a_2 \phi^{+}_2 ,\phi_1 \phi^{+}_2 ,\phi_2 \phi^{+}_1 ,a_1 \phi^{+}_2 ,a_2 \phi^{+}_1 \right\} $.
This matrix can be decomposed as 
\begin{equation}
\mathcal{M}^{ch}_{8\times8}=\left(\begin{array}{cc}
\left(\mathcal{M}^{ch}_1 \right)_{4\times4} & 0\\
0 & \left(\mathcal{M}^{ch}_2 \right)_{4\times4}
\end{array}\right),
\end{equation}
with the first matrix is defined in the basis $\left\{\phi_1 \phi^{+}_1 ,\phi_2 \phi^{+}_2 ,a_1 \phi^{+}_1 ,a_2 \phi^{+}_2 \right\} $
as 
\begin{align}
\mathcal{M}^{ch}_1  & =\left(\begin{array}{cccc}
\frac{\lambda_1}{3} & \frac{\left(\lambda_{4}+\lambda_{5}\right)}{2} & 0 & \frac{i\left(-\lambda_{4}+\lambda_{5}\right)}{2}\\
\frac{\left(\lambda_{4}+\lambda_{5}\right)}{2} & \frac{\lambda_2}{3} & \frac{i\left(-\lambda_{4}+\lambda_{5}\right)}{2} & 0\\
0 & \frac{i\left(\lambda_{4}-\lambda_{5}\right)}{2} & \frac{\lambda_1}{3} & \frac{\left(\lambda_{4}+\lambda_{5}\right)}{2}\\
\frac{i\left(\lambda_{4}-\lambda_{5}\right)}{2} & 0 & \frac{\left(\lambda_{4}+\lambda_{5}\right)}{2} & \frac{\lambda_2}{3}
\end{array}\right),
\end{align}
which leads to the eigenvalues $\frac{\lambda_1 +\lambda_2}{6}\pm\sqrt{\left(\frac{\lambda_1 -\lambda_2}{6}\right)^{2}+\lambda^{2}_{4}}$
and $\frac{\lambda_1 +\lambda_2}{6}\pm\sqrt{\left(\frac{\lambda_1 -\lambda_2}{6}\right)^{2}+\lambda^{2}_{5}}$.
The second submatrix is defined in the basis $\left\{\phi_1 \phi^{+}_2 ,\phi_2 \phi^{+}_1 ,a_1 \phi^{+}_2 ,a_2 \phi^{+}_1 \right\} $
as 
\begin{align}
\mathcal{M}^{ch}_2  & =\left(\begin{array}{cccc}
\lambda_{3} & \frac{\left(\lambda_{4}+\lambda_{5}\right)}{2} & 0 & \frac{i\left(\lambda_{4}-\lambda_{5}\right)}{2}\\
\frac{\left(\lambda_{4}+\lambda_{5}\right)}{2} & \lambda_{3} & \frac{i\left(\lambda_{4}-\lambda_{5}\right)}{2} & 0\\
0 & \frac{i\left(-\lambda_{4}+\lambda_{5}\right)}{2} & \lambda_{3} & \frac{\left(\lambda_{4}+\lambda_{5}\right)}{2}\\
\frac{i\left(-\lambda_{4}+\lambda_{5}\right)}{2} & 0 & \frac{\left(\lambda_{4}+\lambda_{5}\right)}{2} & \lambda_{3}
\end{array}\right),
\end{align}
that gives the eigenvalues $\lambda_{3}\pm\lambda_{4}$ and $\lambda_{3}\pm\lambda_{5}$.

\end{document}